\documentclass{aa}
\usepackage{float}
\usepackage{graphicx}
\usepackage{txfonts}
\usepackage[normalem]{ulem}
\usepackage{xparse}
\usepackage{expl3}

\usepackage{amsmath}
\usepackage{upgreek}
\usepackage[dvipsnames]{xcolor}
\usepackage{hyperref}
\hypersetup{
    colorlinks = true,
    linkcolor = blue,
    anchorcolor = black,
    citecolor = blue,
    filecolor = cyan,
    menucolor = red,
    runcolor = cyan,
    urlcolor = purple
}
\usepackage{diagbox}
\usepackage{siunitx}

\newcommand{\orcid}[1]{%
  \href{https://orcid.org/#1}
}

\usepackage[flushmargin]{footmisc}

\begin{document}

\title{Centrally concentrated star formation in young clusters II:\\ Jet feedback}

\author{
Adilkhan Assilkhan\inst{1,2,3}\thanks{Corresponding author. \email{adilkhan.assilkhan@nu.edu.kz}} \and
Sabrina M. Appel \inst{4} \thanks{NSF Astronomy \& Astrophysics Postdoctoral Fellow} \and
Bekdaulet Shukirgaliyev\inst{5,1,2} \thanks{ \email{bekdaulet.shukirgaliyev@nu.edu.kz}} \and
Ernazar Abdikamalov\inst{1,5} \and
Simon {Portegies Zwart}\inst{6} \and
Eric P. Andersson \inst{4} \and
Mukhagali Kalambay\inst{2,7,8} \and
Mordecai-Mark {Mac Low}\inst{4}
}

\institute{
Energetic Cosmos Laboratory, Nazarbayev University, 53 Kabanbay Batyr Ave., Astana, 010000, Kazakhstan
\and
K. Zhubanov Aktobe Regional University, 34 Moldagulova ave., Aktobe, 030000, Kazakhstan
\and
Gumarbek Daukeyev, Almaty University of Power Engineering and Telecommunications, 126/1 Baytursynuli St., Almaty, 050000, Kazakhstan
\and
Department of Astrophysics, American Museum of Natural History, 200 Central Park W., New York, NY, 10024, USA
\and
Physics Department, Nazarbayev University, 53 Kabanbay Batyr Ave., Astana, 010000, Kazakhstan
\and
Sterrewacht Leiden, Leiden University, Leiden, Netherlands
\and 
Fesenkov Astrophysical Institute, 23 Observatory str., 050020 Almaty, Kazakhstan
\and
Faculty of Physics and Technology, Al-Farabi Kazakh National University, al-Farabi ave. 71, 050040 Almaty, Kazakhstan
}

\date{Received XXXX XX, 2026}

\abstract{
    Protostellar jets are one of the earliest forms of stellar feedback, but their impact on star formation and cluster assembly in centrally concentrated molecular clouds remains poorly understood. We study how protostellar jets affect the star formation efficiency, the temporal variability of star formation, star cluster structure, and the early dynamical state of centrally concentrated, newly forming star clusters using the \texttt{Torch} star-cluster formation framework. We adopt a centrally concentrated initial cloud model with mass $M = 2.5\times 10^3 M_\odot$ and compare six pairs of simulations with and without protostellar jets, supplemented by one additional higher-resolution pair of simulations. We analyze our simulations using global star formation diagnostics together with structural and dynamical measures of the stellar population. Models with jet feedback achieve star formation efficiencies of 12–16\%, while the corresponding models without jets yield higher efficiencies of 19–33\%. Jets also cause star formation to occur in discrete bursts rather than continuously, to produce more extended and substructured stellar systems, and to leave behind stellar populations that are less tightly bound and have higher virial parameters. In our centrally concentrated initial conditions, runs with jets form stellar systems that better reproduce the observed range of the projected structural parameter \(Q_{\mathrm{2D}}\) in young clusters than runs without jets, indicating that protostellar jets are an important early feedback channel even in centrally concentrated clouds that regulates star formation efficiencies and shapes the emerging cluster structure.
}
\keywords{(Galaxy:) open clusters and associations: general --
Stars: formation --
Magnetohydrodynamics (MHD) --
ISM: jets and outflows --
Stars: kinematics and dynamics}
\maketitle
    
\section{Introduction}
\label{sec:intro}

Stars form in giant molecular clouds and more often in embedded clustered environments rather than in isolation \citep{2003ARA&A..41...57L,bressert2010,2019ARA&A..57..227K}. The formation of a young star cluster may be viewed as a sequence in which gas first undergoes gravitational collapse, then forms stars while remaining embedded, and is finally dispersed or restructured by stellar feedback \citep{Krumholz2014a,2020SSRv..216...50C}. The timing and efficiency of this feedback are important because it regulates both the star formation efficiency (SFE) and the likelihood that the resulting stellar system will survive as a bound cluster after gas removal \citep{Geyer2001,Baumgardt2007,portegies-zwart2010, bek+2019, Marina+2025}. Furthermore, determining whether newly formed stellar systems remain bound is important for understanding how the present-day observed cluster population emerges \citep{Bissekenov2024, bissekenov2024exploring, Kalambay2025, Kalambay2026}.

Despite major progress \citep{2021MNRAS.506.3239G,polak2024}, the early feedback phase remains a topic of investigation. In particular, different feedback channels become important at different stages of cluster assembly, and their relative influence depends on when the relevant stellar sources appear. Massive stars can strongly affect the surrounding gas through ionizing radiation and winds, but the formation of massive stars itself remains poorly understood, especially in the context of clustered collapse and the earliest embedded stages \citep{2009MNRAS.400.1775S, Haugbolle+2018A,2018ARA&A..56...41M,2023ApJ...944..211L}. 

\defcitealias{2026A&A...705A..79A}{Paper~I}
In \citet[][hereafter \citetalias{2026A&A...705A..79A}]{2026A&A...705A..79A}, we began to investigate feedback in centrally concentrated gas clouds and showed that such initial conditions can drive efficient central cluster assembly in the presence of stellar feedback from main sequence stars. These initial conditions provide an idealized and controlled experiment in which the shortest free-fall times occur in the inner regions, focusing collapse toward a common central reservoir. This makes them a useful framework for isolating how different feedback channels modify cluster formation in a system initially biased toward rapid central assembly.

However, before massive-star feedback becomes dominant, star formation can already be influenced by an earlier and more widespread feedback channel. Protostellar jets and outflows are launched by accreting young stellar objects and therefore represent one of the earliest forms of stellar feedback. By injecting momentum into the surrounding gas, jets can excavate cavities, redirect inflowing gas, stir turbulence, and reduce the amount of material available for further accretion \citep{2014prpl.conf..451F,2016ARA&A..54..491B}. Numerical studies have shown that protostellar outflows can lower star formation efficiencies and alter the morphology and kinematics of star-forming gas, although the magnitude and even the qualitative character of the effect can depend on the cloud structure, feedback combination, spatial scale, and numerical resolution \citep{2021MNRAS.502.3646G,2022MNRAS.515.4929G,2022A&A...663A...6V,Sabrina2022, Sabrina2023,2024A&A...683A..13L}. 

It remains unclear how protostellar jets affect cluster assembly in centrally concentrated clouds; here we address this question, and build on \citetalias{2026A&A...705A..79A}, by considering simulations with and without jets while using the same numerical framework. Jet feedback in \texttt{Torch} has been implemented by \citet{Sabrina2025}, who described initial comparisons of models with and without jets. Therefore, we keep the cloud density distribution from \citetalias{2026A&A...705A..79A} fixed and only add protostellar jets as the new physical ingredient. To ensure our results are robust, we explore two numerical resolutions and three stochastic realizations, labeled consistently with \citetalias{2026A&A...705A..79A}, as well as an additional realization at a third, higher resolution. The different realizations differ in the seed for the NumPy random-number sequence used by \texttt{Torch} for initial mass function (IMF) sampling and perturbations of initial star-particle positions, velocities, and jet directions. This set of runs allows us to isolate the effects of jets on the SFE, the time history of star formation, the emerging cluster structure, and the early dynamical state of the newly-forming star cluster, all within a well-defined centrally concentrated setup. We also use the projected structural parameter \(Q_{\mathrm{2D}}\) \citep{cartwright2009measuring} to compare the simulations with observed young clusters and to test whether similar projected \(Q_{\mathrm{2D}}\) values correspond to similar stellar morphologies.

The paper is organized as follows: Sect.~\ref{sec:methods} describes our methods and diagnostics; Sect.~\ref{sec:results} presents the effects of protostellar jets on cluster evolution; and Sect.~\ref{sec:conclusions} summarizes our conclusions.

\section{Methods}
\label{sec:methods}

\subsection{Numerical framework}

We study clustered star formation with the same numerical framework as in \citetalias{2026A&A...705A..79A} but with a new simulation suite that includes protostellar jet feedback as implemented by \citet{Sabrina2025}. Our simulations are performed with the \texttt{Torch} framework \citep{wall2019,wall2020}, which uses the Astrophysical Multi-Purpose Software Environment  \citep[\texttt{AMUSE};][]{PortegiesZwart2009,2013A&A...557A..84P,PortegiesZwart2013} and the \texttt{FLASH} code \citep{fryxell2000,dubey2014} to simulate the formation and evolution of star clusters. The gas is evolved with \texttt{FLASH} 4.6.2 \citep{fryxell2000,dubey2014}, while stellar dynamics is followed with the fourth-order Hermite \(N\)-body integrator \texttt{ph4} \citep{mcmillan2012}. Stellar evolution is treated with \texttt{SeBa} \citep{PortegiesZwartVerbunt1996,2012A&A...546A..70T}.

The coupling between the gas and stellar components is handled with a BRIDGE-type Hamiltonian splitting scheme \citep{fujii2007,2020CNSNS..8505240P}. Gas self-gravity, including the contribution from sink particles, is solved on the adaptive mesh, while stellar particles interact gravitationally via direct summation with the \texttt{ph4} $N$-body integrator and through their coupling to the gas potential. For the MHD solver, we adopt the HLLD Riemann solver \citep{2005JCoPh.208..315M} together with piecewise parabolic reconstruction \citep{1984JCoPh..54..174C}.

In \citetalias{2026A&A...705A..79A}, we used \href{https://bitbucket.org/torch-sf/torch/src/torch-v1.0/}{\texttt{Torch version 1.0}}\footnote{\url{https://bitbucket.org/torch-sf/torch/src/torch-v1.0/}, commit b87d873} together with \texttt{AMUSE} commit \texttt{044ca1f7a}.
In the present work, we use the modified \texttt{Torch} version described by \citet{Sabrina2025}, namely the commit tagged \href{https://bitbucket.org/torch-sf/torch/src/jets-v1.2/}{\texttt{jets-v1.2}},\footnote{\url{https://bitbucket.org/torch-sf/torch/src/jets-v1.2/}, commit \texttt{1aceb49}} together with \texttt{AMUSE} commit \texttt{5dff550a4}. This version includes the protostellar jets feedback method described in Sect.~\ref{sect:jets} below.

\subsection{Sink particles and star formation}
\label{sec:star_form}

As in \citetalias{2026A&A...705A..79A}, unresolved gravitational collapse is modelled with sink particles using the \texttt{FLASH} implementation of \citet{federrath2010}. We choose the sink formation threshold density \(\rho_{\rm sink}\) and sink accretion radius \(r_{\rm sink}\) such that the local Jeans length corresponding to \(\rho_{\rm sink}\) satisfies
\begin{equation}
\lambda_{\rm J} = 6\,\Delta x_{\rm min} = 2\,r_{\rm sink},
\end{equation}
where \(\Delta x_{\rm min}\) denotes the minimum cell size on the finest AMR level. Thus, the Jeans length is resolved by six cells during sink formation, in accordance with the requirement in \citep{truelove1997} that the Jeans length be resolved by at least 4 grid cells to avoid artificial fragmentation.
This corresponds to a sink formation threshold density of
\begin{equation}
    \rho_{\mathrm{sink}}
    = \frac{\pi c_s^2}{G\lambda_{\rm J}^2}
    = \frac{\pi c_s^2}{G(6\Delta x_{\rm min})^2}
    = \frac{\pi c_s^2}{4G r_{\rm sink}^2},
    \label{eq:rho_sink}
\end{equation}
where \(G\) is the gravitational constant and \(c_s\) is the isothermal sound speed \citep{truelove1997,federrath2010}. A sink particle is created only if the sink density threshold is exceeded and the additional boundedness and converging-flow criteria of \citet{federrath2010} are satisfied.

After formation, each sink accretes any gas that is above the sink threshold density and within the sink's accretion radius. On formation, each sink is assigned a list of stellar masses drawn by Poisson sampling from the \citet{kroupa2002} IMF over the mass range $0.08$--$150\,M_\odot$ \citep{2017MNRAS.466..407S,wall2019}. A star particle is created only if the sink has accreted enough gas mass to form the next star in its assigned list; that mass is then transferred from the sink particle to the star particle, ensuring the total mass budget is conserved. No primordial binaries are initialized in the present models, consistent with \citetalias{2026A&A...705A..79A}.

\subsection{Initial conditions}

\begin{figure}
  \centering
  \includegraphics[width=\columnwidth]{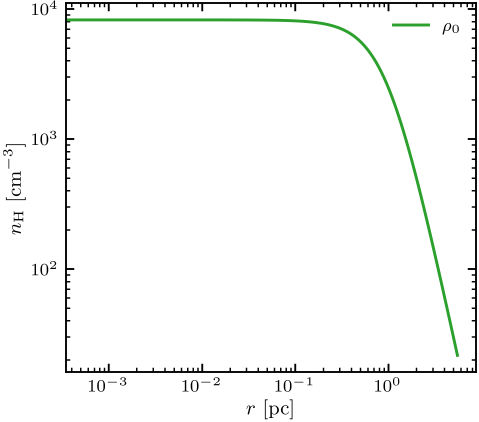}
  \caption{Initial centrally concentrated gas density profile used in our simulations, shown as the hydrogen number density $n_\mathrm{H}(r)$ versus radius $r$ in parsecs. The profile is identical to the initial gas profile in \citetalias{2026A&A...705A..79A} (their Fig.~1), and we adopt the same (green) colour for direct visual comparison.}
  \label{fig:rho0_profile}
\end{figure}

The initial conditions are based on the centrally concentrated molecular cloud model of \citetalias{2026A&A...705A..79A}. The gas density profile is constructed from the analytic centrally concentrated star formation scenario of \citet{2013A&A...549A.132P}, as modified by \citet{Bek+2017}, by requiring that the total (gas + stars) density profile remains fixed during an idealized star-formation episode. In this construction, the stellar component is assumed to follow a Plummer profile but is used only to define the initial gas distribution. The simulations are initialized with gas only, and stars form later from sink particles according to the prescription described in Sect.~\ref{sec:star_form}. For a detailed derivation of the profile, we refer the reader to Sect.~2.5 of \citetalias{2026A&A...705A..79A}. The resulting initial gas density profile, expressed as the hydrogen number density $n_\mathrm{H}(r)$, is shown in Fig.~\ref{fig:rho0_profile}. This initial profile is identical to the one used in \citetalias{2026A&A...705A..79A}, as we employ the same initial data file in both studies.

The cloud has a radius
\begin{equation}
    R_{\rm cl}=5.5\,\mathrm{pc}=5a_\star,
\end{equation}
with \(a_\star=1.1\,\mathrm{pc}\). The initial cloud gas mass is \(M_{\rm cl}=2513\,M_\odot\). The cloud is embedded in a low-density ambient medium, so that the total gas mass in the computational domain is \(3597\,M_\odot\). The computational box extends from \(-6.875\) to \(+6.875\,\mathrm{pc}\) in each spatial direction, corresponding to a box side length of \(L_{\rm box}=13.75\,\mathrm{pc}\).

The gas is initially isothermal at \(T_{\rm gas,0}=20\,\mathrm{K}\) in both the cloud and the ambient medium. Within \(r\leq R_{\rm cl}\), the cloud is seeded with a turbulent velocity field generated using the turbulent-sphere initial-condition procedure of \citet{2015HiA....16..614W}. The turbulent energy spectrum is Kolmogorov-like \citep{1941DoSSR..30..301K},
\begin{equation}
    E(k) \propto k^{-5/3},
\end{equation}
over the wavenumber range \(1\leq k\leq 32\). The surrounding ambient gas is initially static. The turbulent velocity field inside the cloud is rescaled such that the initial gas virial parameter is
\begin{equation}
    \alpha_{\rm vir,gas,0}=0.5.
\end{equation}
The turbulent velocity field is generated by a separate initial-condition procedure with its own random seed. It is independent of the NumPy seeds used by \texttt{Torch} during the star formation calculation and is kept identical across all models.

A weak uniform magnetic field is imposed in the vertical direction,
\begin{equation}
    \mathbf{B}_0=(0,0,3.0)\,\mu{\rm G},
\end{equation}
as in \citetalias{2026A&A...705A..79A}. We use the standard \texttt{FLASH} outflow hydrodynamic boundary conditions, which permit gas inflow from ghost cells, while gravity is treated with isolated boundary conditions.

The adopted global properties of the initial cloud and computational domain are summarized in Table~\ref{tab:cloud_params}. These parameters are fixed for all models considered here. Thus, for a given resolution and random seed, differences between the paired models with jets and without jets can be substantially attributed to the inclusion of protostellar jets. Differences between realizations do reflect further sources of randomness.  These include the sampling and perturbations controlled by the random seed in \texttt{Torch}, as well as the intrinsic run-to-run variability from numerical diffusion in a chaotic flow; this persists even between runs with identical seeds and physics. The additional resolution levels test the numerical robustness of the trends.

\begin{table}
\caption{Initial cloud and computational-domain properties for all models.}
\label{tab:cloud_params}
\centering
\begin{tabular}{lc}
\hline\hline
Parameter & Value \\
\hline
Box side length & \(L_{\rm box}=13.75\,\mathrm{pc}\) \\
Computational domain & \([-6.875,6.875]^3\,\mathrm{pc}^3\) \\
Total gas mass in box & \(3597\,M_\odot\) \\
Cloud gas mass & \(2513\,M_\odot\) \\
Cloud radius & \(R_{\rm cl}=5.5\,\mathrm{pc}\) \\
Plummer radius used in density model & \(a_\star=1.1\,\mathrm{pc}\) \\
Initial gas temperature & \(T_{\rm gas,0}=20\,\mathrm{K}\) \\
Initial gas virial parameter & \(\alpha_{\rm vir,gas,0}=0.5\) \\
Turbulent spectrum & \(E(k)\propto k^{-5/3}\) \\
Turbulent wavenumbers & \(1\leq k\leq 32\) \\
Initial magnetic field strength & \(B_0=3.0\,\mu{\rm G}\) \\
Initial magnetic field direction & \(+\hat{z}\) \\
\hline
\end{tabular} 
\end{table}

\subsection{Resolution levels}

We generally restrict our analysis to runs with two different maximum refinement levels, $n=3$ and $n=4$, corresponding to maximum effective resolutions of $64^3$ and $128^3$ zones and in contrast to \citetalias{2026A&A...705A..79A}, where a broader range of refinement levels was explored. The corresponding minimum cell sizes are $\Delta x_{\rm min}=0.214\,\mathrm{pc}$ for $n=3$ and $\Delta x_{\rm min}=0.107\,\mathrm{pc}$ for $n=4$. The associated sink accretion radii are $r_{\rm sink}=0.64\,\mathrm{pc}$ and $0.32\,\mathrm{pc}$, and the sink threshold densities are $\rho_{\rm sink}=1065\,\mathrm{cm}^{-3}$ and $4262\,\mathrm{cm}^{-3}$.

In addition, we do include one pair higher-resolution realizations as a targeted resolution check motivated by the convergence behavior found in \citetalias{2026A&A...705A..79A}. This pair has \(n=5\), corresponding to an effective resolution of \(256^3\) zones, \(\Delta x_{\rm min}=0.053\,\mathrm{pc}\), \(r_{\rm sink}=0.16\,\mathrm{pc}\), and $\rho_{\rm sink}=17049\,\mathrm{cm}^{-3}$. Because the \(n=5\) calculations are substantially more computationally expensive, only one realization was evolved at this resolution, and one of the simulations did not reach the same final time as our other runs. We therefore only use this pair as a consistency check and not as part of the main ensemble of simulations.

\subsection{Protostellar model with jets}
\label{sect:jets}

The only new physics relative to \citetalias{2026A&A...705A..79A} is the inclusion of protostellar jets. We adopt the jet prescription implemented in \texttt{Torch} by \citet{Sabrina2025}, which adapts the protostellar outflow models described by \citet{Cunningham2011}. We use the default parameters from \citet{Sabrina2025}, as described in Table~\ref{tab:jet_params}.

Jets are launched from star particles with masses in the range
$1\,M_\odot < M_{\star} < 7\,M_\odot$ for a fixed duration at the beginning of the star's main sequence evolution.
For our runs the jet lifetime is $t_{\rm jet} = 100\,\mathrm{kyr}$.
For each eligible star particle, a mass of gas corresponding to a fraction of the star's mass is injected at a constant rate for the duration of the jet.
The total mass injected by each jet is described by
$M_{\rm jet} = f_{\rm mass}\, M_{\star}$, where \(f_{\rm mass} = 0.33\) for our runs. The jet velocity is prescribed as
\begin{equation}
v_{\rm jet} = f_{\rm vel}\left(\frac{G M_{\star}}{r_{\star}}\right)^{1/2},
\end{equation}
where \(r_{\star}\) is the main sequence stellar radius from SeBa \citep{2012A&A...546A..70T} and \(f_{\rm vel} = 0.25\) parametrizes the difference in radius between the main sequence stellar surface and the jet-launching radius at the inner edge of an accretion disk around a protostar. The jet is injected along the star particle's angular-momentum axis with the angular and radial dependence described in \citet{Sabrina2025}\footnote{Note that we use $\Delta\theta = \arctan(1/8)$ in the expression for the angular dependence, as described in Appendix C of \citet{Sabrina2025}.}, including a small random perturbation in the jet direction to model subgrid physics that is not resolved.
 
\begin{table}
\caption{Parameters of the protostellar jet module.}
\label{tab:jet_params}
\centering
\begin{tabular}{lcl}
\hline\hline
Parameter & Value & Description \\
\hline
\(M_{\rm min}\) & \(1.0\,M_\odot\) & Minimum stellar mass for jet injection \\
\(M_{\rm max}\) & \(7.0\,M_\odot\) & Maximum stellar mass for jet injection \\
\(t_{\rm jet}\) & \(100\,\mathrm{kyr}\) & Jet injection duration \\
\(f_{\rm mass}\) & \(0.33\) & Injected mass fraction \\
\(f_{\rm vel}\) & \(0.25\) & Jet velocity fraction \\
\hline
\end{tabular}
\tablefoot{
The parameter \(f_{\rm mass}\) is the injected mass fraction in units of the stellar mass, while \(f_{\rm vel}\) is the jet velocity fraction in units of the Keplerian velocity.
}
\end{table}

\subsection{Simulations}

We present a main suite of 12 simulations, together with one additional higher-resolution pair used as a targeted resolution check. Our simulation suite is summarized in Table~\ref{tab:simulations}. For maximum refinement levels \(n=3\) and \(n=4\), we consider three stochastic realizations labeled \(s=1,3,\) and \(8\). In \citetalias{2026A&A...705A..79A}, ten stochastic realizations at \(n=3\) were evolved and numbered sequentially from 1 to 10; the present simulations reuse three of those realizations (\(s=1\), \(3\), and \(8\)), retaining the original labels to facilitate direct comparison. The corresponding seeds for the NumPy random number generator, specified by \texttt{p['npy\_seed']}, are 0, 2, and 10 for \(s=1\), \(3\), and \(8\), respectively. These seeds initialize the random number sequence used by the Python codes in \texttt{Torch} and therefore determine the stochastic sampling of stellar masses from the \citet{kroupa2002} IMF, as well as the random perturbations applied to the star particle positions, velocities, and jet directions relative to the sink particle properties. Each combination of refinement level and random seed is evolved once without protostellar jets (NoJ) and once with protostellar jets (JET). This paired design allows us to isolate the effect of jet feedback at fixed numerical resolution and fixed random seed, while the multiple realizations allow exploration of the variation associated with the random seed used for both IMF sampling and random perturbations applied to star particle and jet properties.

Our naming convention specifies the maximum refinement level, the realization label, and whether the run includes jets. The additional \texttt{n5s8} pair follows the same qualitative trends as the lower-resolution pairs of simulations in the diagnostics where it is shown. Because \texttt{n5s8-NoJ} covers a shorter evolutionary interval than \texttt{n5s8-JET}, we include this pair only in diagnostics where a direct comparison over the common time interval is informative, and we do not use it in ensemble median quantities.

\begin{table}
\caption{Summary of the simulation suite.}
\label{tab:simulations}
\centering
\small
\begin{tabular}{lcccc}
\hline\hline
Name & \(n\) & \(\Delta x_{\rm min}\) [pc] & Jets & Realization \\
\hline
\texttt{n3s1-NoJ} & 3 & 0.214 & No  & 1 \\
\texttt{n3s3-NoJ} & 3 & 0.214 & No  & 3 \\
\texttt{n3s8-NoJ} & 3 & 0.214 & No  & 8 \\
\texttt{n3s1-JET} & 3 & 0.214 & Yes & 1 \\
\texttt{n3s3-JET} & 3 & 0.214 & Yes & 3 \\
\texttt{n3s8-JET} & 3 & 0.214 & Yes & 8 \\
\hline
\texttt{n4s1-NoJ} & 4 & 0.107 & No  & 1 \\
\texttt{n4s3-NoJ} & 4 & 0.107 & No  & 3 \\
\texttt{n4s8-NoJ} & 4 & 0.107 & No  & 8 \\
\texttt{n4s1-JET} & 4 & 0.107 & Yes & 1 \\
\texttt{n4s3-JET} & 4 & 0.107 & Yes & 3 \\
\texttt{n4s8-JET} & 4 & 0.107 & Yes & 8 \\
\hline
\texttt{n5s8-NoJ} & 5 & 0.054 & No  & 8 \\
\texttt{n5s8-JET} & 5 & 0.054 & Yes & 8 \\
\hline
\end{tabular}
\tablefoot{
\textit{Name} gives the simulation identifier; \(n\) is the maximum AMR refinement level; \(\Delta x_{\rm min}\) is the minimum cell size; \textit{Jets} indicates whether protostellar jet feedback is included; and \textit{Realization} gives the label \(s\) inherited from \citetalias{2026A&A...705A..79A}. The naming convention is \texttt{nXsY-ZZZ}, where \texttt{X} denotes the refinement level, \texttt{Y} denotes the realization label, and \texttt{ZZZ} is either \texttt{JET} or \texttt{NoJ}.
}
\end{table}

\subsection{Cluster structure diagnostics}
\label{subsec:structure_methods}

To quantify the morphology of the stellar distribution, we use three diagnostics: the structural parameter \(Q\) \citep{cartwright2004statistical} evaluated in both 2D projections and 3D space, the number of subclusters \(N_{\rm cl}\), and the stellar half-mass radius \(r_{\rm h}\). The \(Q\) parameter provides a compact measure of stellar spatial structure by comparing the connectivity of the stellar distribution to its characteristic extent. It is defined as \citep{cartwright2004statistical}
\begin{equation}
    Q=\bar{m}/\bar{s},
    \label{eq:q_parameter}
\end{equation}
where \(\bar{m}\) is the normalized mean edge length of the minimum spanning tree and \(\bar{s}\) is the normalized mean inter-particle separation. We compute these quantities as
\begin{equation}
    \bar{m}
    =
    \begin{cases}
    \displaystyle
    \frac{m}{(A/N_\star)^{1/2}}, & \mathrm{2D}, \\[1.5ex]
    \displaystyle
    \frac{m}{(V/N_\star)^{1/3}}, & \mathrm{3D},
    \end{cases}
    \label{eq:mbar_q}
\end{equation}
and
\begin{equation}
    \bar{s}=s/R,
    \label{eq:sbar_q}
\end{equation}
where \(m\) is the mean edge length of the minimum spanning tree, \(s\) is the mean inter-particle separation, \(N_\star\) is the number of stars, \(R\) is the characteristic radius of the stellar distribution, \(A=\pi R^2\) is the projected area in 2D, and \(V=4\pi R^3/3\) is the volume in 3D. Lower values of \(Q\) indicate a more substructured or fractal distribution, whereas larger values correspond to smoother and more centrally concentrated morphologies. Following previous work, we adopt \(Q_{\mathrm{3D}} < 0.7\) as the empirical criterion for fractality in 3D \citep{cartwright2009measuring,2026arXiv260316183A}, and \(Q_{\mathrm{2D}}<0.8\) in projection \citep{cartwright2004statistical}.

For each snapshot, we determine the stellar center of mass and evaluate \(Q_{\mathrm{2D}}\) and \(Q_{\mathrm{3D}}\) for the stars enclosed within the radius containing 97\% of the total stellar mass. This reduces the influence of a small number of distant outliers while capturing the global morphology of the main stellar system. We evaluate \(Q_{\mathrm{2D}}\) in the three projected planes \((x,y)\), \((x,z)\), and \((y,z)\), and adopt the mean value of \(Q_{\mathrm{2D}}\) over the three projections. We use the projected diagnostic to compare to observed young clusters, and to assess whether similar projected \(Q_\mathrm{2D}\) values correspond to similar stellar morphologies.

We also estimate the number of subclusters \(N_{\rm cl}\) with a stellar density-based clustering analysis. For snapshots containing at least 65 stars, we use the \texttt{HOPHaloFinder} implementation in \texttt{yt\_astro\_analysis}, based on the HOP group-finding method \citep{1998ApJ...498..137E,2010ApJS..191...43S}. HOP estimates a local density around each star and links stars by iteratively moving from lower-density particles to their densest neighbours until a local density maximum is reached. We adopt a dimensionless HOP overdensity threshold of 20, which determines which density-linked groups are retained as candidate stellar groups. We then count only groups containing at least 10 stars as subclusters; the resulting number of such groups is recorded as \(N_{\rm cl}\). 

For snapshots with fewer than 65 stars, we instead apply a fallback HOP-like algorithm directly to the stellar positions and masses. Local stellar densities are estimated from \(k\)-nearest neighbours, with \(k=\min(32,\,N_\ast-1)\), and each star is iteratively linked to the densest neighbour until convergence to a local density maximum, yielding initial peak-centred groups \citep{1998ApJ...498..137E}. Only stars with \(\rho>0.5\,\bar{\rho}\) are considered for grouping, and only peaks with \(\rho_{\rm peak}>\bar{\rho}\) are retained. Nearby candidate groups are then merged when their estimated saddle density exceeds \(0.8\,\min(\rho_{{\rm peak},1},\rho_{{\rm peak},2})\). Finally, as in the main HOP analysis, only groups with at least 10 members are kept in the final count. As an additional size diagnostic, we compute the stellar half-mass radius \(r_{\rm h}\), defined as the 3D radius around the stellar center of mass enclosing 50\% of the total stellar mass.

To quantify the agreement between simulations and observations, we compute the root-mean-square (RMS) difference between the median simulation curves and the unweighted binned median relation of the observed \(Q_{\mathrm{2D}}\) as a function of age, evaluated on a common age grid. The observed median relation is constructed by binning the observational data in age and computing the median \(Q_{\mathrm{2D}}\) value in each bin. We further assess the robustness of this comparison using bootstrap resampling of the observational sample \citep{efron1979bootstrap}, in which the observed clusters are resampled with replacement and the binned median relation is recomputed for each realization.

\subsection{Dynamical diagnostics}
\label{subsec:dynamical_methods}

To characterize the early dynamical state of the stellar population, we compute the mass fractions of weakly bound stars \(M_{\rm b,wk}/M_\star\) and of strongly bound stars \(M_{\rm b,str}/M_\star\), as well as the stellar virial ratio \(\alpha_{\rm vir,\star}\). These quantities are used to assess how tightly bound and how close to virial equilibrium the stellar population is \citep{2012MNRAS.419..841K,2015MNRAS.450.2451F,Bek+2017,bek+2019,2019MNRAS.487..364L}.
We consider a star $i$ to be weakly bound if its kinetic energy \(K_i\) satisfies
\begin{equation}
K_i + m_i \phi_{\star,i} < 0,
\end{equation}
where \(m_i\) is the stellar mass, and \(\phi_{\star,i}\) is the gravitational potential generated by the other stars.
We consider stars to be strongly bound if they are bound by more than twice their kinetic energy, such that
\begin{equation}
2K_i + m_i \phi_{\star,i} < 0.
\end{equation}
Note that the mass fraction of weakly bound stars includes all strongly bound stars.
We evaluate these thresholds by including both the stellar potential and the gas potential \(m_i[\Phi(\mathbf{x}_i)-\Phi_{\rm ref}]\), where the constant reference potential \(\Phi_{\rm ref}\) is subtracted to account for the arbitrary zero-point of the gas potential.

The stellar virial ratio of the cluster is defined by
\begin{equation}
\alpha_{\rm vir,\star} = 2K/|W_\star|,
\label{eq:alphavir}
\end{equation}
where \(K\) is the total stellar kinetic energy and
\begin{equation}
W_\star = \frac{1}{2}\sum_i m_i \phi_{\star,i}
\end{equation}
is the stellar self-gravitational energy. Thus, the weakly and strongly bound fractions are computed using both the stellar and gas gravitational potentials, whereas \(\alpha_{\rm vir,\star}\) is computed using only the stellar self-gravity. Values \(\alpha_{\rm vir,\star}\approx 1\) indicate approximate virial equilibrium of the stars alone, whereas larger values correspond to dynamically hotter and less tightly bound stellar systems. Since cluster survival after gas removal is sensitive to the early dynamical state, these diagnostics are important for understanding cluster survival \citep{2003ARA&A..41...57L, bek+2018, Bek+2021,  
2021MNRAS.508.5410D}.

\section{Results} \label{sec:results}

\subsection{Impact on cluster structure}
\label{subsec:stellar_structure}

\begin{figure}[!htbp] 
  \centering
  \includegraphics[width=\columnwidth]{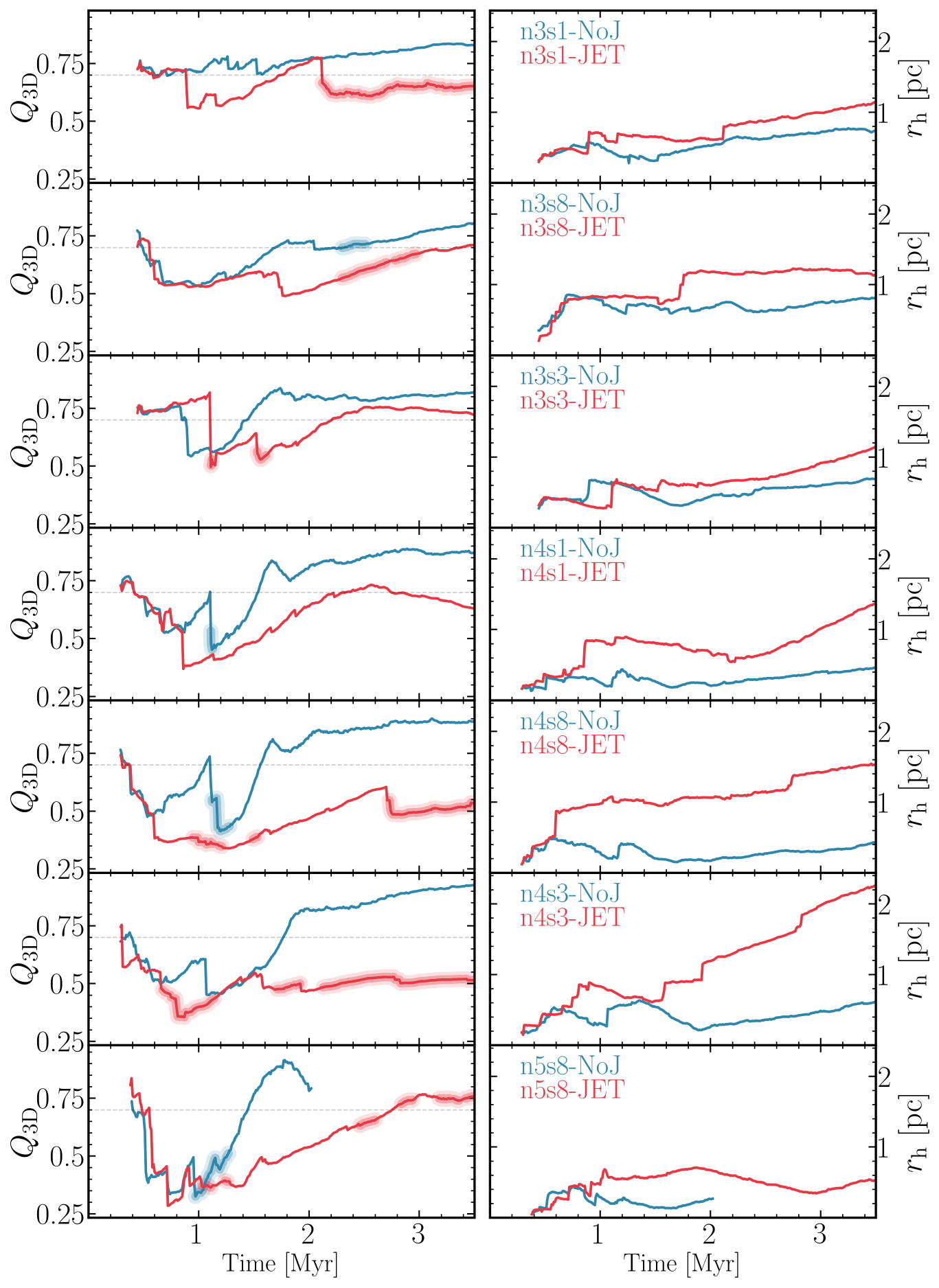}
  \caption{
  Time evolution of the stellar structural diagnostics for the models with (JET; red curves) and without (NoJ; blue curves) jets. The left column shows the structural parameter \(Q_{\mathrm{3D}}\) and the right column shows the stellar half-mass radius \(r_{\rm h}\). The rows show, from top to bottom, the models \texttt{n3s1}, \texttt{n3s8}, \texttt{n3s3}, \texttt{n4s1}, \texttt{n4s8}, \texttt{n4s3}, and the additional higher-resolution check \texttt{n5s8}. The horizontal dashed line in the left panels marks \(Q_{\mathrm{3D}} = 0.7\), the threshold below which clusters are more substructured and above which clusters are smoother and more centrally concentrated. Shaded segments in the \(Q_{\mathrm{3D}}\) panels indicate epochs when the clustering analysis identifies more than one subcluster (\(N_{\rm cl} > 1\)).
  }
  \label{fig:structure_qrh}
\end{figure}

First, we examine how protostellar jets impact the structure of newly forming stellar systems using the cluster structure diagnostics defined in Sect.~\ref{subsec:structure_methods}: \(Q_{\mathrm{3D}}\), \(N_{\rm cl}\), and \(r_{\rm h}\). Here, \(Q_{\mathrm{3D}}\) provides a continuous measure of the degree of substructure in the stellar distribution, \(N_{\rm cl}\) identifies epochs with multiple distinct subclusters, and \(r_{\rm h}\) measures the overall spatial size of the stellar system.

Figure~\ref{fig:structure_qrh} shows that the models without jets (labelled NoJ) generally evolve toward larger \(Q_{\mathrm{3D}}\) values, eventually exceeding the adopted transition threshold and thus indicating a smoother, more centrally concentrated \emph{stellar} morphology. As discussed in \citetalias{2026A&A...705A..79A}, centrally concentrated clouds have the shortest free-fall times in their inner regions, so global collapse drives gas toward the center. In the NoJ runs this leads both to the formation of new stars at smaller radii and to the inward motion and merging of initially distinct subclusters, which together erase stellar substructure and increase \(Q_{\mathrm{3D}}\). By contrast, the models with jets (labelled JET) spend substantially more time at lower \(Q_{\mathrm{3D}}\); several of the JET runs remain below or close to the transition threshold of 0.7 for most of their evolution. The lower values of \(Q_{\mathrm{3D}}\) and the occasional drops in \(Q_{\mathrm{3D}}\) are often associated with epochs when \(N_{\rm cl}>1\), as indicated by the shaded segments in the left column of Fig.~\ref{fig:structure_qrh}. This indicates that both an increase in the number of subclusters and a more extended or dispersed spatial distribution of stars can drive \(Q_{\mathrm{3D}}\) to lower values. The low and irregular values of \(Q_{\mathrm{3D}}\) in the JET runs therefore suggest that protostellar jet feedback promotes subclustering and irregular, non-monolithic stellar distributions.

The evolution of \(r_{\rm h}\), shown in the right panels of Fig.~\ref{fig:structure_qrh}, supports this interpretation. In the main \(n=3,4\) suite, the models with jets generally have larger half-mass radii at late times than their counterparts without jets, indicating that the stellar distribution is usually more extended when jets are included. Importantly, the lower \(Q_{\mathrm{3D}}\) values in the jets runs are not solely due to the presence of sharply separated subclusters. In some realizations (such as the \verb|n4s1| and \verb|n4s8| runs), \(Q_{\mathrm{3D}}\) remains low and \(r_{\rm h}\) stays comparatively large even while the clustering analysis identifies only one subcluster. This indicates that jet feedback can produce irregular stellar distributions in two ways: by splitting the stellar distribution into multiple subclusters, or by producing a broader, more diffuse stellar distribution. The additional \texttt{n5s8} pair shows that individual high-resolution realizations can differ from the size trend seen in the main \(n=3,4\) suite. In this pair, both the NoJ and JET runs have systematically smaller half-mass radii than most of the lower-resolution realizations, even though \texttt{n5s8-JET} still maintains a larger \(r_{\rm h}\) than \texttt{n5s8-NoJ} at all times. We interpret this as a consequence of weaker stellar feedback in these runs: in \texttt{n5s8-JET}, no star more massive than \(18\,M_\odot\) forms during the simulated evolution, so the feedback history and the redistribution of dense gas differ from those in several of the lower-resolution realizations that do form more massive stars. This supports the picture that the structural response to protostellar jets is modulated by stochastic sampling of the high-mass end of the stellar population, rather than being set by numerical resolution alone, and complements earlier results showing that the cluster response to jet feedback is sensitive to feedback from early-forming massive stars.

\begin{figure*}[!htbp]
  \centering
  \includegraphics[width=0.9\textwidth]{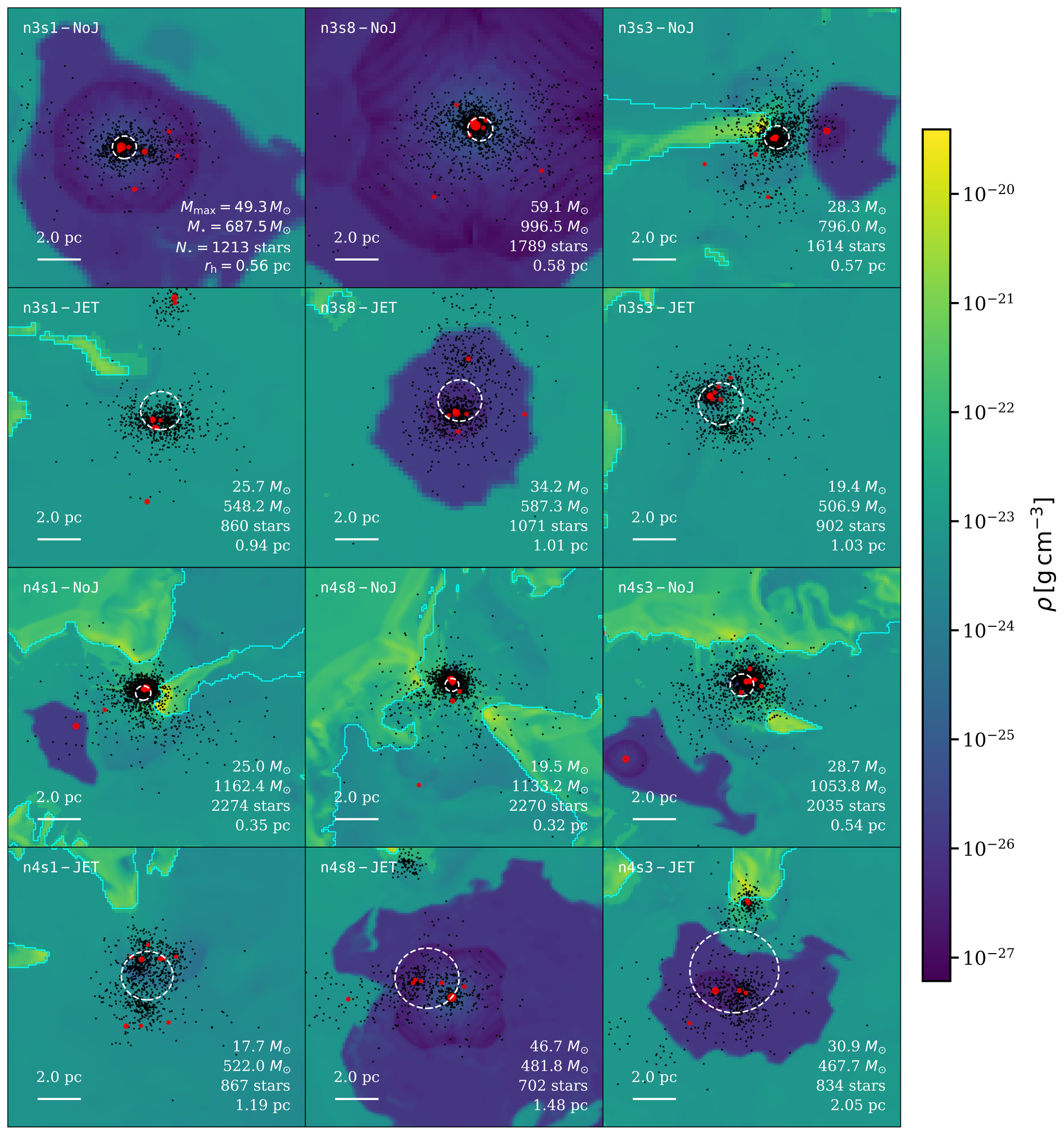}
  \caption{
  Gas density slices in the \(x\)--\(y\) plane at \(t \simeq 3.5\) Myr for the main \(n=3,4\) simulation suite. In each panel, the slice is centered on the \(z\)-position of the stellar density peak. {\em Black dots} show all stars below \(7\,M_\odot\). {\em Red circles} indicate massive stars with \(M_\star \geq 7\,M_\odot\), where the radius of each circle is proportional to the star's mass. The {\em dashed white circle} marks the projected stellar half-mass radius \(r_{\rm h}\) in the \(x\)--\(y\) plane relative to the projected stellar center of mass. \textit{Cyan contours} trace the boundary of the ionized gas. The model name is shown in the top-left corner of each panel, while the mass of the most massive star \(M_{\rm max}\), the total stellar mass \(M_{\star}\), the number of stars \(N_{\star}\), and the half-mass radius \(r_{\rm h}\) are shown in the bottom-right corner. A white bar in the lower-left corner of each panel indicates a scale of 2 pc.
  }
  \label{fig:density_jets}
\end{figure*}

The structural differences implied by Fig.~\ref{fig:structure_qrh} are visible in the snapshots shown in Fig.~\ref{fig:density_jets}. The slices in Fig.~\ref{fig:density_jets} are centered on the peak of the stellar density and thus enable a comparison of the morphology and degree of concentration of the cluster center. The projected stellar half-mass radius \(r_{\rm h}\) is also measured with respect to the projected stellar center of mass in the \(x\)-\(y\) plane and provides a consistent measure of the size of the stellar distribution.

In the runs without jets, the stellar population is typically concentrated around a dominant central gas peak and the projected half-mass radius remains comparatively small. In the runs with jets, the gas distribution is more strongly perturbed, low-density cavities are more prominent, and the stellar component is more spatially extended. Several runs with jets exhibit either multiple local condensations or a broader asymmetric stellar distribution, consistent with the lower \(Q_{\mathrm{3D}}\) values and larger \(r_{\rm h}\) found in Fig.~\ref{fig:structure_qrh}. The structural diagnostics and the morphology are mutually consistent: protostellar jets hinder the formation of a single compact stellar system, instead favoring a more extended, and in some cases more substructured, mode of cluster assembly.

A plausible physical explanation is that jet feedback acts at the stage when the first generation of stars begins to assemble in the central collapsing region. Since jets are launched by newly formed low- and intermediate-mass protostars, they inject momentum early, while the dense central gas is still collapsing inward. In our centrally concentrated clouds, this does not simply shut off star formation everywhere. Instead, it perturbs the central collapsing reservoir and displaces part of the dense gas from the main collapse zone, sometimes redirecting part of the subsequent collapse to nearby sites. Depending on the specific realization, this can produce either (temporarily) distinct subclusters or a single but more extended stellar system. Jet feedback modifies not only the efficiency of star formation, but also the geometry of stellar assembly.

This comparison also shows that the structural impact of protostellar jets is not yet fully understood and can depend on initial conditions. In the massive-clump simulations of \citet{2022A&A...663A...6V}, which adopt a more extended initial gas distribution than our runs, the inclusion of jets and \ion{H}{II} regions increases the global \(Q\) parameter and reduces or delays the occurrence of long-lived substructures in the cluster outskirts, so that the stellar configuration becomes more centrally concentrated overall. By contrast, driven turbulent-box calculations show that protostellar jets can enhance small-scale fragmentation, producing a larger number of sink particles than otherwise comparable magnetized runs \citep{Sabrina2023}, even though the large-scale gas distribution remains comparatively extended.

Resolution differences alone are unlikely to explain these contrasting outcomes. Our previous centrally concentrated simulations without jets show that, at fixed initial density profile, increasing the maximum resolution tends to accelerate central cluster assembly and produce a more compact stellar concentration rather than enhancing long-lived substructure (see Fig.~2 in \citetalias{2026A&A...705A..79A}). At the same time, the number of sink particles generally increases with resolution in all of these studies, so a higher sink count does not by itself guarantee a more strongly subclustered large-scale stellar configuration. Thus, differences in the initial gas profile and in the physical scale of the simulated region can be as important as resolution when comparing the effects of jets across simulations. In practice, the measured impact of jets also depends on the adopted sink and feedback prescriptions, the spatial scale on which the stellar structure is analyzed, and whether one focuses on the number of sink particles, the global \(Q\) parameter, or the number of spatially distinct stellar subclusters as the primary diagnostic.

Our simulations probe a different regime from the massive-clump models of \citet{2022A&A...663A...6V} and driven turbulent-box calculations, namely we consider a centrally concentrated cloud in which the shortest free-fall times occur near the center and collapse is initially focused toward a common dense reservoir \citepalias{2026A&A...705A..79A}. In this geometry, protostellar jets do not primarily remove pre-existing off-center condensations. Instead, they perturb the central star-forming gas, reduce and intermittently redirect accretion, and can displace dense material away from the original collapse center. As a result, the stellar distribution in our jet runs is often more extended and, in many realizations, more strongly subclustered than in the corresponding runs without jets, as reflected in the systematically larger \(N_{\rm cl}\) and lower \(Q_{\mathrm{3D}}\). We therefore interpret the enhanced substructure in our models as the combined outcome of jet-driven fragmentation and redistribution of the central gas reservoir within this centrally concentrated setup, rather than as evidence for any single universal tendency of protostellar jets to either increase or decrease substructure in all environments.

\subsubsection{Projected structure comparison}
\label{subsec:q2d_obs}

Having established the structural differences between the runs with and without jets from our 3D diagnostics, we now examine the projected \(Q_{\mathrm{2D}}\) parameter. We use this parameter for two purposes: first, to compare our simulations with observed young clusters; and second, to assess whether similar projected \(Q_\mathrm{2D}\) values correspond to similar projected cluster morphologies.

\begin{figure}[!htbp]
  \centering
  \includegraphics[width=\columnwidth]{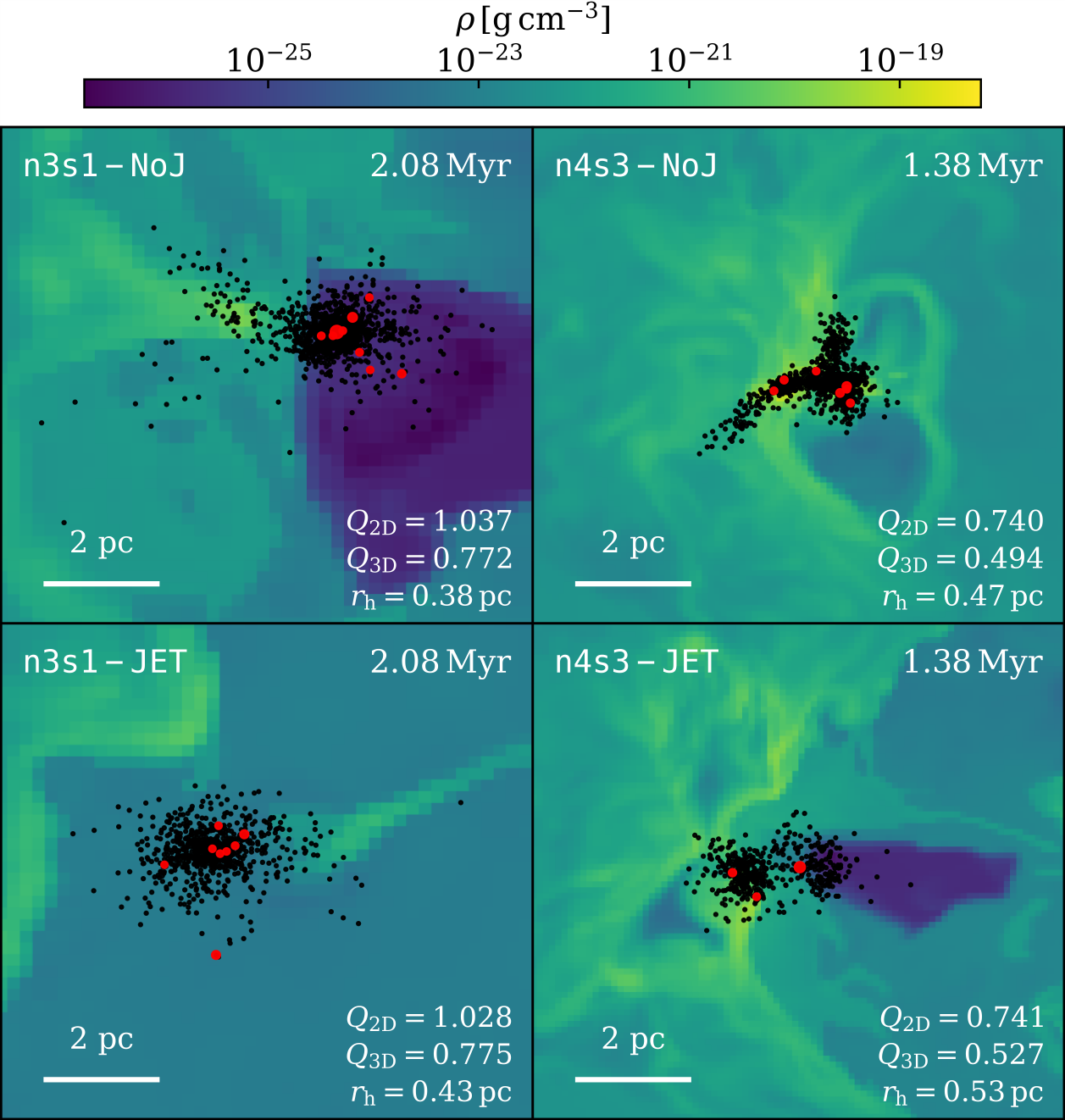}
  \caption{
  Gas density slices for two pairs of models with (JET) and without (NoJ) jets selected at epochs when the projected structural parameter \(Q_{\mathrm{2D}}\) is nearly equal within each pair. The left column shows the \texttt{n3s1} model pair at \(2.08\) Myr, and the right column shows the \texttt{n4s3} pair at \(1.38\) Myr; the top row shows the NoJ models and the bottom row shows the JET models. The upper color bar shows the gas density, \(\rho\), in units of \(\mathrm{g\,cm^{-3}}\).
  {\em Black dots} show all stellar particles, while {\em red circles} highlight massive stars with \(M_\star \geq 7\,M_\odot\), with the marker size scaled by stellar mass. The model name is shown in the top-left corner of each panel, the simulation time is shown in the top-right corner, and \(Q_{\mathrm{2D}}\), \(Q_{\mathrm{3D}}\), and the projected half-mass radius \(r_{\rm h}\) are shown in the lower-right corner. A white bar in the lower-left corner indicates a scale of 2 pc.
  }
  \label{fig:q2d_matched}
\end{figure}

Figure~\ref{fig:q2d_matched} illustrates how the morphological interpretation of the projected \(Q_{\mathrm{2D}}\) parameter depends on the underlying structure of the stellar distribution. For the \texttt{n3s1} and \texttt{n4s3} pairs, the figure shows the central values of \(Q_{\mathrm{2D}}\), \(Q_{\mathrm{3D}}\), and \(r_{\rm h}\), along with gas density slices at corresponding times for each model pair. For the \texttt{n3s1} pair, the runs without and with jets have similar projected \(Q_{\mathrm{2D}}\) values, with uncertainties computed using a leave-one-out jackknife method: \(Q_{\mathrm{2D}} = 1.037 \pm 0.024\) and \(1.028 \pm 0.018\), respectively. Their 3D structural parameters are likewise very similar: \(Q_{\mathrm{3D}} = 0.772 \pm 0.016\) and \(0.775 \pm 0.018\). In both \texttt{n3s1} panels, the stellar distribution is dominated by a single main concentration rather than by clearly separated subclusters. Although the detailed stellar distributions and asymmetries differ, these examples suggest that in the smoother regime (\(Q_{\mathrm{2D}}>0.8\)), similar projected \(Q_{\mathrm{2D}}\) values can be associated with broadly similar large-scale projected morphologies.

However, the \texttt{n4s3} pair of runs illustrates a different situation. Here, the two models again have nearly equal \(Q_{\mathrm{2D}}\) values within the jackknife uncertainties: \(Q_{\mathrm{2D}} = 0.740 \pm 0.032\) and \(0.741 \pm 0.022\). In addition, their 3D structural parameters remain well within the substructured regime: \(Q_{\mathrm{3D}} = 0.494 \pm 0.025\) and \(0.527 \pm 0.017\). However, examining the right column of Fig.~\ref{fig:q2d_matched} clearly indicates that the projected stellar distributions are not morphologically identical. The model without jets appears by eye to be more filamentary and irregular, whereas the model with jets is more compact but still markedly asymmetric, with a clear extension away from the main concentration. These two runs demonstrate that similar \(Q_{\mathrm{2D}}\) values do not uniquely describe the qualitative projected morphology when the stellar distribution remains strongly substructured.

We suggest that the interpretive power of the projected structural parameter \(Q_{\mathrm{2D}}\) is limited to determining the presence of substructure, rather than its specific type. For stellar distributions that lack significant substructure, similar \(Q_{\mathrm{2D}}\) values can correspond to qualitatively similar morphologies. However, highly substructured clusters may be substructured in different ways, as seen in the right column of Fig.~\ref{fig:q2d_matched}. In such cases, similar projected \(Q_{\mathrm{2D}}\) values may correspond to different morphologies, since \(Q_{\mathrm{2D}}\) measures the overall degree of spatial irregularity. Thus, \(Q_{\mathrm{2D}}\) is a useful statistical indicator of substructure, but it should not be over-interpreted as a unique classifier of specific morphologies.

\subsubsection{Comparison to observed values}
\label{subsec:q2d_obs_pt2}

\begin{figure*}[!htbp]
  \centering
  \includegraphics[width=0.75\linewidth]{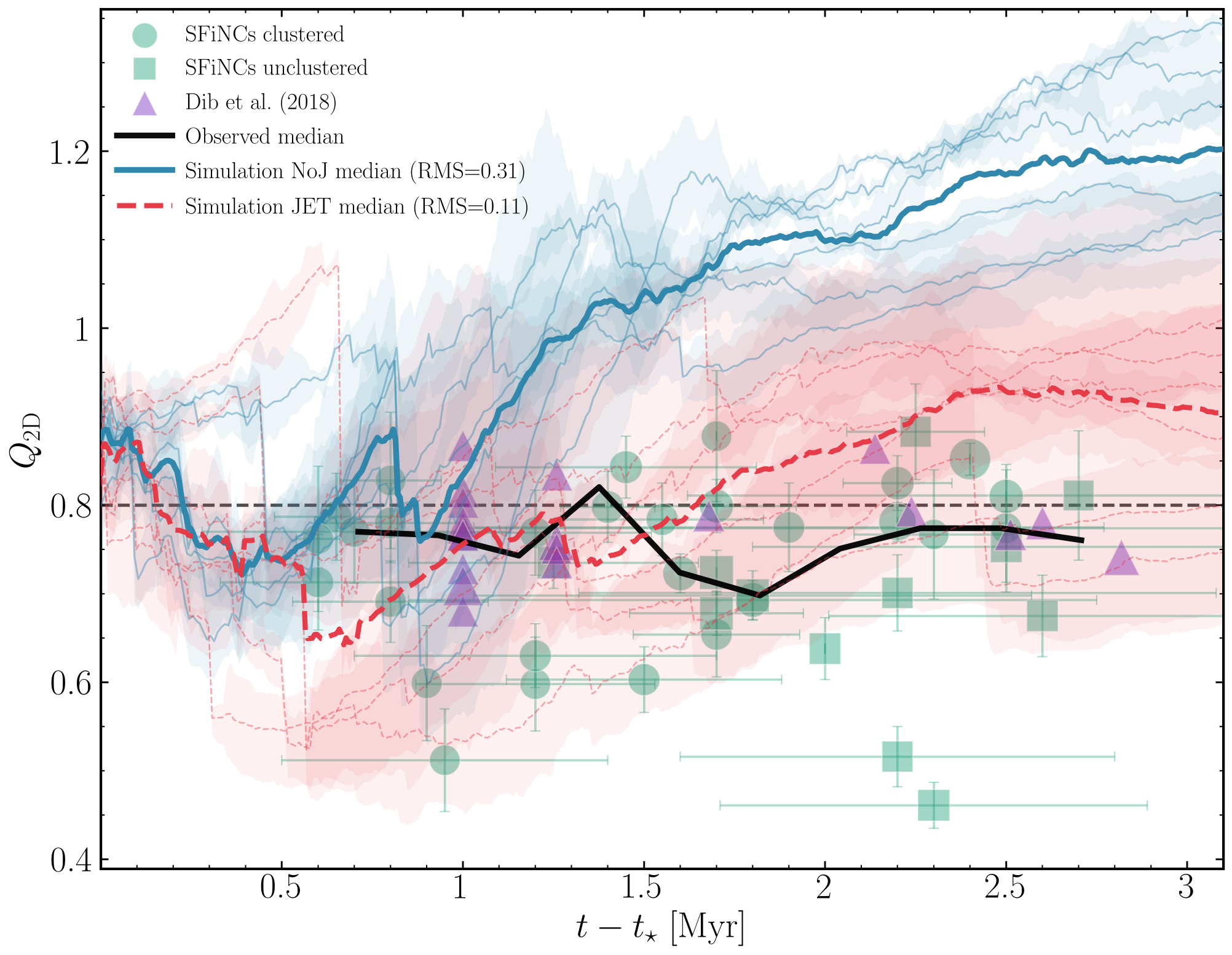}
  \caption{
  Evolution of the projected structural parameter \(Q_{\mathrm{2D}}\) as a function of stellar age measured from the onset of star formation, \(t-t_\star\). Thin red dashed and blue solid  curves show the mean value of the three line-of-sight projections for individual runs with and without jets. The thick curves show the median evolution of each group of models. The shaded regions around the individual simulation tracks show the full minimum–maximum range of \(Q_{\mathrm{2D}}\) values over the three projection axes. Observational data are included for comparison: green circles denote clustered \textit{SFiNCs} systems, green squares denote the extended unclustered \textit{SFiNCs} components \citep{getman2018}, and purple triangles show the young open-cluster sample from \citet{dib2018}. For \textit{SFiNCs}, horizontal and vertical error bars show the jackknife \(1\sigma\) uncertainties in age and \(Q_{\mathrm{2D}}\), respectively. The black curve gives the unweighted binned median of the combined observational sample, computed from the central age and \(Q_{\mathrm{2D}}\) values only; the observational uncertainties are shown for reference but are not used as weights in constructing this median. The legend gives the RMS value of the differences over time between the median simulation curves (with and without jets) and the observed median. The horizontal dashed line marks \(Q_{\mathrm{2D}}=0.8\), below which projected stellar distributions are interpreted as more fractal or substructured.
  }
  \label{fig:q2d_obs}
\end{figure*}

Figure~\ref{fig:q2d_obs} compares the time evolution of the projected structural parameter \(Q_{\mathrm{2D}}\) in observed young stellar systems with our simulations. The observational sample is described in Appendix~\ref{app:obs_sample}. It combines clustered and extended unclustered components from the \textit{SFiNCs} catalogue \citep{getman2018} with the young open-cluster sample analysed by \citet{dib2018}. For \textit{SFiNCs}, we compute \(Q_{\mathrm{2D}}\) from the published stellar positions and estimate jackknife \(1\sigma\) uncertainties for both \(Q_{\mathrm{2D}}\) and the median age. For the open-cluster subsample, we adopt the published structural parameter \(Q\) from \citet{dib2018} and the cluster ages from the MWSC catalogue \citep{mwsc1}. Since no individual age or \(Q\) uncertainties are available for the \citet{dib2018} subsample, these points are treated as point estimates. Only systems younger than \(3.1\) Myr are shown, matching the age range covered by our simulation comparison.

The observational data occupy a broad but well-defined range, with most systems lying near \(Q_{\mathrm{2D}}\sim 0.7\)--0.9. The unweighted binned median of the combined observational sample remains close to \(Q_{\mathrm{2D}}\simeq 0.8\), the approximate transition between smoother centrally concentrated morphologies and more substructured projected distributions. This indicates that many young stellar systems retain a substantial degree of projected spatial irregularity during the first few megayears.

The median simulation curves show a clear separation between the runs with and without jets. In the runs without jets, \(Q_{\mathrm{2D}}\) increases after \(\sim 1\) Myr, rising significantly above the observational median. This rise shows that these systems become smoother and more centrally concentrated in projection than the observed young systems. In contrast, the runs with jets remain at systematically lower \(Q_{\mathrm{2D}}\) values and stay closer to the observed median relation over most of the plotted age range. This qualitative description agrees with the RMS values shown in the legend: the RMS difference between the median simulation curve and the unweighted binned median of the observed \(Q_{\mathrm{2D}}\)--age relation is \(0.105\) for the runs with jets and \(0.308\) for the runs without jets. As a robustness check, we apply bootstrap resampling of the observational sample and recompute this RMS diagnostic for \(10^4\) realizations. In all valid bootstrap realizations, the runs with jets give the smaller RMS, corresponding to \(P_{\rm boot}({\rm jets\ closer})=1.000\) and \(P_{\rm boot}({\rm no\ jets\ closer})<10^{-4}\). We interpret this as a bootstrap measure of relative support for the two simulation sets, rather than as a formal probability that the observed clusters are drawn from either model distribution.

This comparison supports the interpretation developed from the 3D structural diagnostics in Sect.~\ref{subsec:stellar_structure}. Jet feedback maintains a more irregular and extended stellar component, and this remains visible in projection through lower \(Q_{\mathrm{2D}}\) values. The runs without jets can begin within the observed range, but they evolve more rapidly toward smoother projected morphologies. The runs with jets therefore provide a better population-level match to the observed degree of projected substructure in nearby young stellar systems. This agreement should not be over-interpreted for individual objects, since, as shown in Sect.~\ref{subsec:q2d_obs}, similar \(Q_{\mathrm{2D}}\) values do not necessarily imply identical projected morphologies when the stellar distribution remains strongly substructured.

\subsection{Impact on star formation}
\label{subsec:sfe_sfr}

We next quantify how protostellar jets affect the global conversion of gas into stars. For this purpose, we use the box-wide integrated SFE,
\begin{equation}
{\rm SFE}_{\rm box}(t)=M_\star(t)/M_{{\rm tot,box}},
\label{eq:SFE_box}
\end{equation}
where \(M_\star(t)\) is the total stellar mass formed by time \(t\), and \(M_{{\rm tot,box}}=3597\,M_\odot\) is the initial gas mass inside the computational domain. 

\begin{figure}[!htbp] 
    \centering
    \includegraphics[width=\columnwidth]{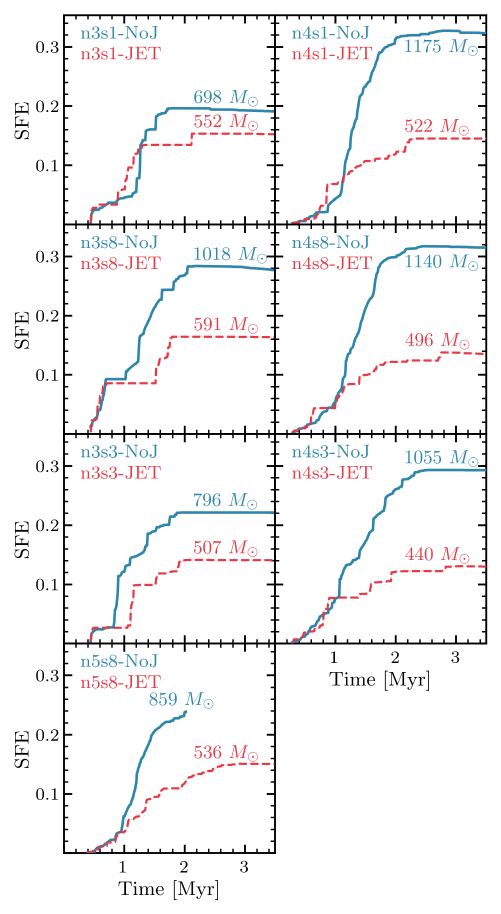}
    \caption{
    Time evolution of the box-integrated SFE (Eq.~\ref{eq:SFE_box}) for the six main NoJ vs. JET simulation pairs and the additional higher-resolution \texttt{n5s8} pair. Each panel compares one simulation without jets (solid blue lines) to the corresponding simulation with jets (red dashed lines). The model names and final stellar masses for each run are shown on each panel. Protostellar jets reduce the final stellar mass and the global SFE.
    }
    \label{fig:sfe}
\end{figure}

\begin{figure}[!htbp]
    \centering
    \includegraphics[width=\columnwidth]{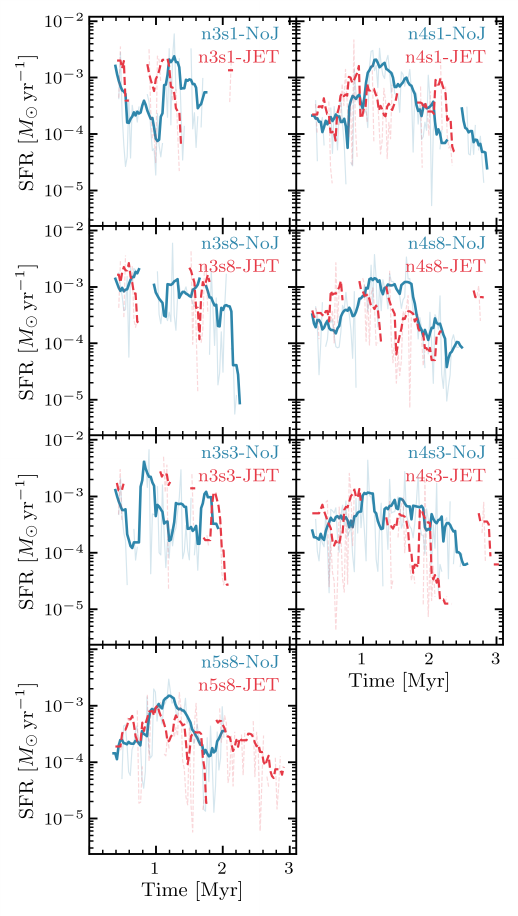}
    \caption{
    Time evolution of the \(\mathrm{SFR}\) for the six main NoJ vs. JET simulation pairs and the additional higher-resolution \texttt{n5s8} pair. Each panel compares one simulation without jets (solid blue lines) to the corresponding simulation with jets (red dashed lines).  The faint curves show the raw binned SFR and the darker curves show the corresponding running-mean trends. The model names are shown on each panel. Overall, the runs with jets tend to display a more irregular, bursty star formation history and, in several cases, a more rapid decline of the SFR at late times. In contrast, the runs without jets tend to continue to form stars over a longer interval. 
    }
    \label{fig:sfr}
\end{figure}

Figure~\ref{fig:sfe} shows that protostellar jets decrease the SFE for each pair in the main \(n=3,4\) suite, and that the additional higher-resolution \texttt{n5s8} pair follows the same qualitative trend over the time interval where the two runs overlap. By the end of the simulations, the models without jets reach ${\rm SFE}_{\rm box}\simeq 0.19$--0.33, whereas the models with jets saturate at only ${\rm SFE}_{\rm box}\simeq 0.12$--0.16. In terms of stellar mass, the models without jets form \(698\)--\(1175\,M_\odot\), while the models with jets produce only \(440\)--\(591\,M_\odot\). The reduction in final stellar mass therefore amounts to a factor of \(\sim 1.3\)--2.4, with the strongest suppression occurring in the higher-resolution \(n4\) models. This trend is consistent with the results of \citet{Sabrina2025}, who used models with flatter initial density profiles but also found that protostellar jets in \texttt{Torch} slow star formation and significantly affect the evolution of lower-mass clouds.

The temporal evolution of the SFE provides additional insight into the feedback mechanism. At early times, the curves of runs with and without jets remain relatively similar, reflecting the fact that that the first stars form before jet feedback has had time to substantially alter the surrounding gas distribution. After roughly 0.7--1 Myr, however, the runs with and without jets diverge. In the runs without jets, the stellar mass continues to increase efficiently as collapse proceeds in the central dense region, and the models ultimately saturate at relatively high values of ${\rm SFE}_{\rm box}$. In contrast, the runs with jets grow more slowly and saturate at systematically lower values of ${\rm SFE}_{\rm box}$. Thus, the main effect of protostellar jets is not simply to cause the SFE to plateau, since all runs eventually approach saturation, but to lower the asymptotic SFE reached by the system. This indicates that the momentum injected by protostellar jets increasingly suppresses subsequent star formation as the runs evolve.

This behavior is apparent in all three stochastic realizations at \(n=3\) and \(n=4\), and is also consistent with the additional \(n=5\), \(s=8\) check. While the exact final efficiencies vary between realizations, this scatter remains much smaller than the systematic separation between the runs with and without jets. The reduced SFE in the models with jets is therefore a general outcome of protostellar jet feedback rather than a feature of any individual realization or due to the choice of resolution.

To further understand the suppression of star formation, we examine the SFR as a function of time for each run. The SFR is reconstructed directly from the stellar particle data by binning the formation times of the star particles, summing the stellar mass formed within each time bin, and dividing by the bin width. In Fig.~\ref{fig:sfr}, we show both the raw binned SFR and a running-mean trend in order to distinguish short-timescale fluctuations from the longer-term evolution.

Figure~\ref{fig:sfr} shows that jets do not only lower the integrated SFE overall; they also change the temporal character of star formation. The runs without jets generally maintain a sustained SFR, with SFRs remaining comparatively continuous, though variable, throughout a phase of active star formation before gradually declining at late times. By contrast, the runs with jets exhibit markedly more intermittent SFRs. Their SFRs are characterized by distinct bursts of activity separated by intervals of strongly reduced or nearly absent star formation, as seen by the gaps in the red curves in Fig.~\ref{fig:sfr}.

Bursty star formation can be understood as a feedback cycle operating in the dense central gas. Newly formed stars inject momentum through jets over a characteristic timescale of \(100\) kyr, excavating low-density cavities and dispersing part of the nearby star-forming material. As a result, local collapse is temporarily weakened, lowering the SFR. Once gas is re-accreted or re-assembled by ongoing global collapse, star formation resumes and the cycle begins again. The runs with jets therefore alternate between active and quiescent phases, with typical interruptions of order \(0.1\)--0.5 Myr. In the runs without jets, this cycle is absent and so collapse proceeds in a more sustained manner.

Taken together, Figs.~\ref{fig:sfe}-\ref{fig:sfr} show that protostellar jets act as an efficient self-regulation mechanism in centrally concentrated clouds. Jets reduce the total amount of gas that is converted into stars and transform star formation from a relatively sustained process into a more episodic one. In our simulations, the dominant effect of jets is therefore not to halt star formation immediately, but to repeatedly disrupt the dense central gas reservoir from which stars form, thereby reducing the overall SFE and resulting in more episodic star formation.

\begin{figure*}[t]
    \centering
    \includegraphics[width=0.9\linewidth]{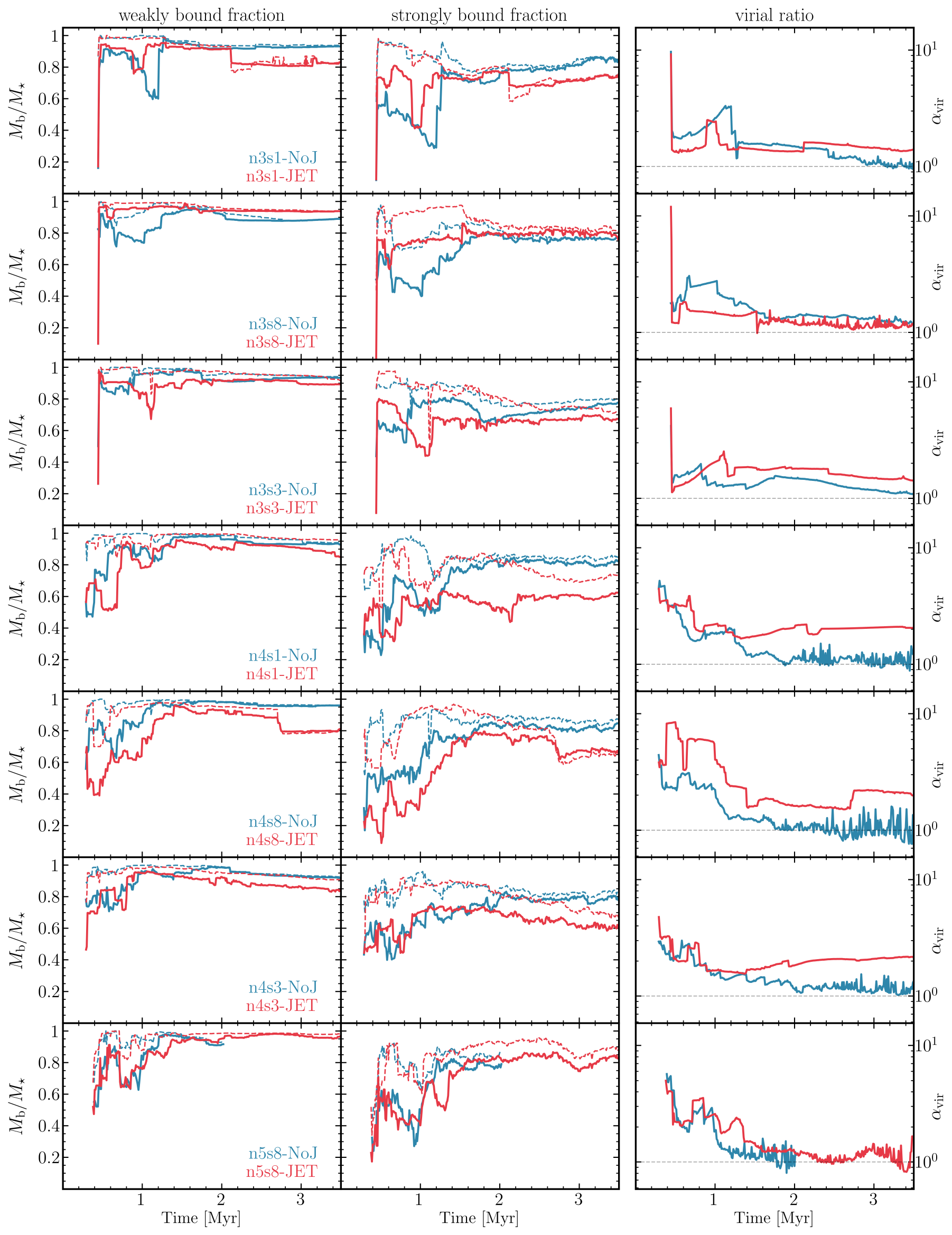}
    \caption{
    Time evolution of the weakly and strongly bound stellar mass fractions, \(M_{\rm b,wk}/M_\star\) and \(M_{\rm b,str}/M_\star\), and of the stellar virial ratio, \(\alpha_{\rm vir}\), for each of our runs. The rows correspond, from top to bottom, to models \texttt{n3s1}, \texttt{n3s8}, \texttt{n3s3}, \texttt{n4s1}, \texttt{n4s8}, \texttt{n4s3}, and the additional higher-resolution model \texttt{n5s8}. {\em Blue curves} denote runs without jets and {\em red curves} denote runs with jets. In the first two columns, {\em solid curves} are computed using the stellar self-gravity only, while {\em dashed curves} additionally include the gas gravitational potential. The {\em left column} shows the weakly bound fraction based on the criterion \(K_i+m_i\phi_{\star,i}<0\), and the {\em middle column} shows the strongly bound fraction based on \(2K_i+m_i\phi_{\star,i}<0\). The {\em right column} shows the stellar virial ratio \(\alpha_{\rm vir,\star}\) (see Eq.~\ref{eq:alphavir}), with a {\em horizontal dashed line} to mark virial equilibrium (\(\alpha_{\rm vir}=1\)).
    }
    \label{fig:bound_virial}
\end{figure*}

\subsection{Impact on stellar dynamics}
\label{subsec:bound_virial}

We next examine how protostellar jets affect the early dynamical state of the stellar distribution using the diagnostics defined in Sect.~\ref{subsec:dynamical_methods}: the weakly and strongly bound stellar mass fractions, \(M_{\rm b,wk}/M_\star\) and \(M_{\rm b,str}/M_\star\), and the stellar virial ratio \(\alpha_{\rm vir}\).

Figure~\ref{fig:bound_virial} shows that protostellar jets generally lead to less strongly bound stellar systems. This trend is clearest in the strongly bound fraction (the middle column), which is a stricter measure of stellar boundedness. In most pairs of runs, the run without jets retains a larger fraction of strongly bound stellar mass at late times, whereas the corresponding run with jets reaches systematically lower values. The same trend is also seen in the weakly bound fraction (the left column), although the differences are smaller there. The one exception is the \texttt{n3s8} pair, in which the run with jets is marginally more bound than the corresponding run without jets. A plausible interpretation is that this reflects the specific star formation and assembly history of these two runs, showing that stochastic differences in star formation can sometimes outweigh the tendency of jet feedback to reduce the bound fraction. The higher-resolution \texttt{n5s8} pair is consistent with this broad dynamical interpretation over the common time interval.

Including the gravitational potential due to the gas (the dashed lines in Fig.~\ref{fig:bound_virial}) increases both the weakly and strongly bound fractions, especially at earlier times. This shows that the remaining gas still contributes to the boundedness of the stellar distribution for several Myr. Nevertheless, the overall trend between runs with and without jets persists. Thus, the lower bound fractions in the models with jets are not simply a consequence of neglecting the gas potential, but reflect a genuine difference in the dynamical state of the stellar system. As the evolution of the runs proceeds, the difference between the fractions computed with and without the gas potential decreases, consistent with the declining dynamical importance of the gas.

The virial ratios support the same interpretation. Several runs without jets, in particular \texttt{n3s1-NoJ}, \texttt{n4s1-NoJ}, and \texttt{n4s8-NoJ}, evolve toward \(\alpha_{\rm vir}\simeq 1\) by the end of the simulations, indicating an approximately virialized state. By contrast, the corresponding runs with jets generally remain more supervirial. The stellar systems formed in the presence of jets are therefore not only less bound, but also dynamically hotter. These dynamical differences parallel the structural trends discussed in Sect.~\ref{subsec:stellar_structure}: in general, runs with jets remain more extended and structurally irregular, tend to retain smaller bound fractions, and have higher virial ratios. This indicates that jet feedback affects not only the spatial distribution of the stars but also their early dynamical state.

These results suggest that protostellar jets affect cluster survivability at very early stages. Because jet feedback operates as soon as stars begin to form, it can weaken the stellar gravitational potential before later feedback channels become dominant. In \citetalias{2026A&A...705A..79A}, we examined the impact of feedback from massive stars once massive stars form. The results of our present paper indicate that protostellar jets can already pre-process the stellar system dynamically by reducing the bound fraction and shifting it toward a more supervirial state. Only a small fraction of embedded clusters are thought to survive as long-lived bound open clusters \citep{2003ARA&A..41...57L,portegies-zwart2010}, and protostellar jet feedback may represent one of the mechanisms contributing to early cluster dispersal. A full assessment of long-term cluster survival, however, will require continuing these simulations into the later gas-poor regime and following the subsequent \(N\)-body evolution, which we defer to future work.

\section{Conclusions}
\label{sec:conclusions}

We have investigated how protostellar jets modify star formation and cluster assembly in centrally concentrated molecular clouds using the same numerical framework as in \citetalias{2026A&A...705A..79A}, but with a new suite of paired simulation runs with and without jets. Our results show that protostellar jets systematically reduce the global SFE and the final stellar mass, and transform star formation from a more sustained process into a more episodic, bursty one. In all pairs of runs, the runs with jets reach lower final SFEs than the corresponding runs without jets, demonstrating that protostellar jets act as an efficient early self-regulation mechanism for these low-mass regions. The higher-resolution \(n=5\), \(s=8\) paired realization supports these qualitative trends where a direct comparison is possible.

Jets also modify the morphology of the stellar distribution. Compared to the runs without jets, the runs with jets remain more substructured, spend more time with a lower structural parameter \(Q_{\mathrm{3D}}\), and generally exhibit larger stellar half-mass radii. Rather than assembling into a single compact central system, the stellar distributions formed with jets are typically more extended and more prone to irregular, extended, or subclustered configurations. In projection, the models with jets better reproduce the observed range of projected structural parameters \(Q_{\mathrm{2D}}\) in young clusters and remain within this range for a larger fraction of their evolution than the corresponding runs without jets.

The early dynamical state of the stellar systems is likewise affected by jet feedback. In most realizations, the runs with jets retain smaller weakly and strongly bound stellar mass fractions and remain more supervirial than their counterparts without jets, indicating that jet feedback generally leaves behind stellar systems that are dynamically hotter and less tightly bound. Taken together, these results suggest that protostellar jets can influence not only how much gas is converted into stars, but also the structure of the resulting star cluster and how likely it is to remain bound at later times.

Within the specific centrally concentrated geometry studied here, protostellar jets are an important ingredient in the early evolution of low-mass star-forming clouds. By reducing the SFE, enhancing the time variability of star formation, and producing stellar systems that are generally more extended and less tightly bound, jets may contribute to the early dispersal of embedded clusters before later feedback from massive stars becomes dominant. At the same time, comparisons with other numerical studies show that the structural response to jet feedback can differ in less concentrated or more extended clouds, and likely depends on the initial cloud concentration and the adopted combination of feedback channels. Taken together with our earlier high-resolution runs without jets, this suggests that the centrally concentrated initial profile and  global collapse scale play key roles in setting how jets redistribute star-forming gas. Furthermore, jets keep the projected structural evolution of the models closer to that of observed young clusters for longer. Following simulations of these systems into the later gas-poor regime will be necessary to determine how strongly these early differences affect the subsequent \(N\)-body evolution and long-term survival of the resulting star clusters.

\section*{Data availability}

Simulation data products used in this study--including time-history tables, stellar particle catalogues, representative raw \texttt{FLASH}/\texttt{Torch} snapshots at two reference epochs, and density-slice movies for all runs--are publicly available via the American Museum of Natural History Digital Library at \href{https://doi.org/10.5531/sd.astro.12} {AMNH}\footnote{\url{https://doi.org/10.5531/sd.astro.12}}. Additional data products, including additional snapshots, are available upon reasonable request.

\section*{Acknowledgments}
 AA acknowledges support from a Bolashaq International Scholarship. The authors thank Nurzhan Ussipov and Almat Akhmetali for fruitful discussions and their extensive support regarding the $Q$ structural parameter. AA, BS and MK acknowledge the support from Grant No.\ AP26102895, funded by the Science Committee of the Ministry of Science and Higher Education of the Republic of Kazakhstan. EA is supported by Nazarbayev University Faculty Development Competitive Research Grant Program (No.\ 040225FD4713). M-MML is partly supported by US National Science Foundation (NSF) grant AST23-07950. EPA and M-MML are supported by NASA ATP grant 80NSSC24K0935. SMA is supported by an NSF Astronomy and Astrophysics Postdoctoral Fellowship under award AST24-01740.
 
 \texttt{FLASH} was developed in part by the DOE NNSA- and DOE Office of Science-supported Flash Center for Computational Science at the University of Chicago and the University of Rochester. Other software used includes \texttt{AMUSE} \citep{PortegiesZwart2009, Pelupessy+2013, PortegiesZwart2013, PortegiesZwartMcMillan2018}, \texttt{SeBa} \citep{PortegiesZwartVerbunt1996,Toonen+2012}, \texttt{ph4} \citep{McMillan+2012, PortegiesZwartMcMillan2018}, \texttt{yt} \citep{Turk11a}, \texttt{SciPy} \citep{SciPy}, \texttt{Matplotlib} \citep{Hunter2007}, and \texttt{astropy} \citep{Astropy-Collaboration13a}. 

Large language models, including ChatGPT and Perplexity, were used during manuscript preparation to assist with grammar, language clarity, and literature exploration. All resulting text and references were carefully reviewed and verified by the authors.

\bibliographystyle{aa}
\bibliography{bib,maclow,sample631}

\begin{thebibliography}{77}
\expandafter\ifx\csname natexlab\endcsname\relax\def\natexlab#1{#1}\fi

\bibitem[{{Akhmetali} {et~al.}(2026){Akhmetali}, {Assilkhan}, {Mac Low}, {Ussipov}, {Zaidyn}, {Abdikamalov}, {Sills}, {Pang}, \& {Shukirgaliyev}}]{2026arXiv260316183A}
{Akhmetali}, A., {Assilkhan}, A., {Mac Low}, M.-M., {et~al.} 2026, arXiv e-prints, arXiv:2603.16183

\bibitem[{{Appel} {et~al.}(2025){Appel}, {Burkhart}, {Mac Low}, {Andersson}, {Cournoyer-Cloutier}, {Lewis}, {McMillan}, {Polak}, {Portegies Zwart}, {Tran}, \& {Wilhelm}}]{Sabrina2025}
{Appel}, S.~M., {Burkhart}, B., {Mac Low}, M.-M., {et~al.} 2025, arXiv e-prints, arXiv:2509.15311

\bibitem[{{Appel} {et~al.}(2022){Appel}, {Burkhart}, {Semenov}, {Federrath}, \& {Rosen}}]{Sabrina2022}
{Appel}, S.~M., {Burkhart}, B., {Semenov}, V.~A., {Federrath}, C., \& {Rosen}, A.~L. 2022, \apj, 927, 75

\bibitem[{{Appel} {et~al.}(2023){Appel}, {Burkhart}, {Semenov}, {Federrath}, {Rosen}, \& {Tan}}]{Sabrina2023}
{Appel}, S.~M., {Burkhart}, B., {Semenov}, V.~A., {et~al.} 2023, \apj, 954, 93

\bibitem[{{Assilkhan} {et~al.}(2026){Assilkhan}, {Mac Low}, {Polak}, {Abdikamalov}, {Cournoyer-Cloutier}, {Lewis}, {Kalambay}, {Otebay}, \& {Shukirgaliyev}}]{2026A&A...705A..79A}
{Assilkhan}, A., {Mac Low}, M.-M., {Polak}, B., {et~al.} 2026, \aap, 705, A79

\bibitem[{{Astropy Collaboration} {et~al.}(2013){Astropy Collaboration}, {Robitaille}, {Tollerud}, {Greenfield}, {Droettboom}, {Bray}, {Aldcroft}, {Davis}, {Ginsburg}, {Price-Whelan}, {Kerzendorf}, {Conley}, {Crighton}, {Barbary}, {Muna}, {Ferguson}, {Grollier}, {Parikh}, {Nair}, {Unther}, {Deil}, {Woillez}, {Conseil}, {Kramer}, {Turner}, {Singer}, {Fox}, {Weaver}, {Zabalza}, {Edwards}, {Azalee Bostroem}, {Burke}, {Casey}, {Crawford}, {Dencheva}, {Ely}, {Jenness}, {Labrie}, {Lim}, {Pierfederici}, {Pontzen}, {Ptak}, {Refsdal}, {Servillat}, \& {Streicher}}]{Astropy-Collaboration13a}
{Astropy Collaboration}, {Robitaille}, T.~P., {Tollerud}, E.~J., {et~al.} 2013, \aap, 558, A33

\bibitem[{{Bally}(2016)}]{2016ARA&A..54..491B}
{Bally}, J. 2016, \araa, 54, 491

\bibitem[{Baumgardt \& Kroupa(2007)}]{Baumgardt2007}
Baumgardt, H. \& Kroupa, P. 2007, MNRAS, 380, 1589

\bibitem[{{Bissekenov} {et~al.}(2024{\natexlab{a}}){Bissekenov}, {Kalambay}, {Abdikamalov}, {Pang}, {Berczik}, \& {Shukirgaliyev}}]{Bissekenov2024}
{Bissekenov}, A., {Kalambay}, M., {Abdikamalov}, E., {et~al.} 2024{\natexlab{a}}, \aap, 689, A282

\bibitem[{{Bissekenov} {et~al.}(2024{\natexlab{b}}){Bissekenov}, Kalambay, Abylkairov, \& Shukirgaliyev}]{bissekenov2024exploring}
{Bissekenov}, A., Kalambay, M.~T., Abylkairov, Y.~S., \& Shukirgaliyev, B.~T. 2024{\natexlab{b}}, Recent Contributions to Physics, 90

\bibitem[{{Bressert} {et~al.}(2010){Bressert}, {Bastian}, {Gutermuth}, {Megeath}, {Allen}, {Evans}, {Rebull}, {Hatchell}, {Johnstone}, {Bourke}, {Cieza}, {Harvey}, {Merin}, {Ray}, \& {Tothill}}]{bressert2010}
{Bressert}, E., {Bastian}, N., {Gutermuth}, R., {et~al.} 2010, \mnras, 409, L54

\bibitem[{Cartwright(2009)}]{cartwright2009measuring}
Cartwright, A. 2009, \mnras, 400, 1427

\bibitem[{Cartwright \& Whitworth(2004)}]{cartwright2004statistical}
Cartwright, A. \& Whitworth, A.~P. 2004, \mnras, 348, 589

\bibitem[{{Chevance} {et~al.}(2020){Chevance}, {Kruijssen}, {Vazquez-Semadeni}, {Nakamura}, {Klessen}, {Ballesteros-Paredes}, {Inutsuka}, {Adamo}, \& {Hennebelle}}]{2020SSRv..216...50C}
{Chevance}, M., {Kruijssen}, J.~M.~D., {Vazquez-Semadeni}, E., {et~al.} 2020, \ssr, 216, 50

\bibitem[{{Colella} \& {Woodward}(1984)}]{1984JCoPh..54..174C}
{Colella}, P. \& {Woodward}, P.~R. 1984, Journal of Computational Physics, 54, 174

\bibitem[{{Cunningham} {et~al.}(2011){Cunningham}, {Klein}, {Krumholz}, \& {McKee}}]{Cunningham2011}
{Cunningham}, A.~J., {Klein}, R.~I., {Krumholz}, M.~R., \& {McKee}, C.~F. 2011, \apj, 740, 107

\bibitem[{{Dib} {et~al.}(2018){Dib}, {Schmeja}, \& {Parker}}]{dib2018}
{Dib}, S., {Schmeja}, S., \& {Parker}, R.~J. 2018, \mnras, 473, 849

\bibitem[{{Dom{\'\i}nguez} {et~al.}(2021){Dom{\'\i}nguez}, {Farias}, {Fellhauer}, \& {Klessen}}]{2021MNRAS.508.5410D}
{Dom{\'\i}nguez}, R., {Farias}, J.~P., {Fellhauer}, M., \& {Klessen}, R.~S. 2021, \mnras, 508, 5410

\bibitem[{Dubey {et~al.}(2014)Dubey, Antypas, Calder, Daley, Fryxell, Gallagher, Lamb, Lee, Olson, Reid, Rich, Ricker, Riley, Rosner, Siegel, Taylor, Weide, Timmes, Vladimirova, \& ZuHone}]{dubey2014}
Dubey, A., Antypas, K., Calder, A.~C., {et~al.} 2014, The International Journal of High Performance Computing Applications, 28, 225

\bibitem[{Efron(1979)}]{efron1979bootstrap}
Efron, B. 1979, The Annals of Statistics, 7, 1

\bibitem[{{Eisenstein} \& {Hut}(1998)}]{1998ApJ...498..137E}
{Eisenstein}, D.~J. \& {Hut}, P. 1998, \apj, 498, 137

\bibitem[{{Farias} {et~al.}(2015){Farias}, {Smith}, {Fellhauer}, {Goodwin}, {Candlish}, {Bla{\~n}a}, \& {Dominguez}}]{2015MNRAS.450.2451F}
{Farias}, J.~P., {Smith}, R., {Fellhauer}, M., {et~al.} 2015, \mnras, 450, 2451

\bibitem[{{Federrath} {et~al.}(2010){Federrath}, {Banerjee}, {Clark}, \& {Klessen}}]{federrath2010}
{Federrath}, C., {Banerjee}, R., {Clark}, P.~C., \& {Klessen}, R.~S. 2010, \apj, 713, 269

\bibitem[{{Frank} {et~al.}(2014){Frank}, {Ray}, {Cabrit}, {Hartigan}, {Arce}, {Bacciotti}, {Bally}, {Benisty}, {Eisl{\"o}ffel}, {G{\"u}del}, {Lebedev}, {Nisini}, \& {Raga}}]{2014prpl.conf..451F}
{Frank}, A., {Ray}, T.~P., {Cabrit}, S., {et~al.} 2014, in Protostars and Planets VI, ed. H.~{Beuther}, R.~S. {Klessen}, C.~P. {Dullemond}, \& T.~{Henning}, 451--474

\bibitem[{{Fryxell} {et~al.}(2000){Fryxell}, {Olson}, {Ricker}, {Timmes}, {Zingale}, {Lamb}, {MacNeice}, {Rosner}, {Truran}, \& {Tufo}}]{fryxell2000}
{Fryxell}, B., {Olson}, K., {Ricker}, P., {et~al.} 2000, \apjs, 131, 273

\bibitem[{{Fujii} {et~al.}(2007){Fujii}, {Iwasawa}, {Funato}, \& {Makino}}]{fujii2007}
{Fujii}, M., {Iwasawa}, M., {Funato}, Y., \& {Makino}, J. 2007, \pasj, 59, 1095

\bibitem[{{Getman} {et~al.}(2017){Getman}, {Broos}, {Kuhn}, {Feigelson}, {Richert}, {Ota}, {Bate}, \& {Garmire}}]{getman2017}
{Getman}, K.~V., {Broos}, P.~S., {Kuhn}, M.~A., {et~al.} 2017, \apjs, 229, 28

\bibitem[{{Getman} {et~al.}(2018){Getman}, {Kuhn}, {Feigelson}, {Broos}, {Bate}, \& {Garmire}}]{getman2018}
{Getman}, K.~V., {Kuhn}, M.~A., {Feigelson}, E.~D., {et~al.} 2018, \mnras, 477, 298

\bibitem[{Geyer \& Burkert(2001)}]{Geyer2001}
Geyer, M.~P. \& Burkert, A. 2001, \mnras, 323, 988

\bibitem[{{Grudi{\'c}} {et~al.}(2021){Grudi{\'c}}, {Kruijssen}, {Faucher-Gigu{\`e}re}, {Hopkins}, {Ma}, {Quataert}, \& {Boylan-Kolchin}}]{2021MNRAS.506.3239G}
{Grudi{\'c}}, M.~Y., {Kruijssen}, J.~M.~D., {Faucher-Gigu{\`e}re}, C.-A., {et~al.} 2021, \mnras, 506, 3239

\bibitem[{{Guszejnov} {et~al.}(2021){Guszejnov}, {Grudi{\'c}}, {Hopkins}, {Offner}, \& {Faucher-Gigu{\`e}re}}]{2021MNRAS.502.3646G}
{Guszejnov}, D., {Grudi{\'c}}, M.~Y., {Hopkins}, P.~F., {Offner}, S. S.~R., \& {Faucher-Gigu{\`e}re}, C.-A. 2021, \mnras, 502, 3646

\bibitem[{{Guszejnov} {et~al.}(2022){Guszejnov}, {Grudi{\'c}}, {Offner}, {Faucher-Gigu{\`e}re}, {Hopkins}, \& {Rosen}}]{2022MNRAS.515.4929G}
{Guszejnov}, D., {Grudi{\'c}}, M.~Y., {Offner}, S. S.~R., {et~al.} 2022, \mnras, 515, 4929

\bibitem[{{Haugb{\o}lle} {et~al.}(2018){Haugb{\o}lle}, {Padoan}, \& {Nordlund}}]{Haugbolle+2018A}
{Haugb{\o}lle}, T., {Padoan}, P., \& {Nordlund}, {\r{A}}. 2018, \apj, 854, 35

\bibitem[{{Hunter}(2007)}]{Hunter2007}
{Hunter}, J.~D. 2007, Computing in Science and Engineering, 9, 90

\bibitem[{{Ishchenko} {et~al.}(2025){Ishchenko}, {Masliukh}, {Hradov}, {Berczik}, {Shukirgaliyev}, \& {Omarov}}]{Marina+2025}
{Ishchenko}, M., {Masliukh}, V., {Hradov}, M., {et~al.} 2025, \aap, 694, A33

\bibitem[{{Kalambay} {et~al.}(2026){Kalambay}, {Ishchenko}, {Kuvatova}, {Panamarev}, \& {Berczik}}]{Kalambay2026}
{Kalambay}, M., {Ishchenko}, M., {Kuvatova}, D., {Panamarev}, T., \& {Berczik}, P. 2026, \aap, 708, A89

\bibitem[{Kalambay {et~al.}(2025)Kalambay, Otebay, Nazar, Assilkhan, \& Shukirgaliyev}]{Kalambay2025}
Kalambay, M.~T., Otebay, A.~B., Nazar, A.~B., Assilkhan, A., \& Shukirgaliyev, B.~T. 2025, Herald of the Kazakh-British Technical University, 22, 312

\bibitem[{{Kharchenko} {et~al.}(2013){Kharchenko}, {Piskunov}, {Schilbach}, {R{\"o}ser}, \& {Scholz}}]{mwsc1}
{Kharchenko}, N.~V., {Piskunov}, A.~E., {Schilbach}, E., {R{\"o}ser}, S., \& {Scholz}, R.-D. 2013, \aap, 558, A53

\bibitem[{{Kolmogorov}(1941)}]{1941DoSSR..30..301K}
{Kolmogorov}, A. 1941, Akademiia Nauk SSSR Doklady, 30, 301

\bibitem[{{Kroupa}(2002)}]{kroupa2002}
{Kroupa}, P. 2002, Science, 295, 82

\bibitem[{{Kruijssen} {et~al.}(2012){Kruijssen}, {Maschberger}, {Moeckel}, {Clarke}, {Bastian}, \& {Bonnell}}]{2012MNRAS.419..841K}
{Kruijssen}, J.~M.~D., {Maschberger}, T., {Moeckel}, N., {et~al.} 2012, \mnras, 419, 841

\bibitem[{{Krumholz} {et~al.}(2014){Krumholz}, {Bate}, {Arce}, {Dale}, {Gutermuth}, {Klein}, {Li}, {Nakamura}, \& {Zhang}}]{Krumholz2014a}
{Krumholz}, M.~R., {Bate}, M.~R., {Arce}, H.~G., {et~al.} 2014, in Protostars and Planets VI, ed. H.~Beuther, R.~S. Klessen, C.~P. Dullemond, \& T.~Henning (Tucson: U. of Arizona Press), 243--266

\bibitem[{{Krumholz} {et~al.}(2019){Krumholz}, {McKee}, \& {Bland-Hawthorn}}]{2019ARA&A..57..227K}
{Krumholz}, M.~R., {McKee}, C.~F., \& {Bland-Hawthorn}, J. 2019, \araa, 57, 227

\bibitem[{{Lada} \& {Lada}(2003)}]{2003ARA&A..41...57L}
{Lada}, C.~J. \& {Lada}, E.~A. 2003, \araa, 41, 57

\bibitem[{{Lebreuilly} {et~al.}(2024){Lebreuilly}, {Hennebelle}, {Maury}, {Gonz{\'a}lez}, {Traficante}, {Klessen}, {Testi}, \& {Molinari}}]{2024A&A...683A..13L}
{Lebreuilly}, U., {Hennebelle}, P., {Maury}, A., {et~al.} 2024, \aap, 683, A13

\bibitem[{{Lewis} {et~al.}(2023){Lewis}, {McMillan}, {Mac Low}, {Cournoyer-Cloutier}, {Polak}, {Wilhelm}, {Tran}, {Sills}, {Portegies Zwart}, {Klessen}, \& {Wall}}]{2023ApJ...944..211L}
{Lewis}, S.~C., {McMillan}, S. L.~W., {Mac Low}, M.-M., {et~al.} 2023, \apj, 944, 211

\bibitem[{{Li} {et~al.}(2019){Li}, {Vogelsberger}, {Marinacci}, \& {Gnedin}}]{2019MNRAS.487..364L}
{Li}, H., {Vogelsberger}, M., {Marinacci}, F., \& {Gnedin}, O.~Y. 2019, \mnras, 487, 364

\bibitem[{{McMillan} {et~al.}(2012{\natexlab{a}}){McMillan}, {Portegies Zwart}, {van Elteren}, \& {Whitehead}}]{mcmillan2012}
{McMillan}, S., {Portegies Zwart}, S., {van Elteren}, A., \& {Whitehead}, A. 2012{\natexlab{a}}, in Astronomical Society of the Pacific Conference Series, Vol. 453, Advances in Computational Astrophysics: Methods, Tools, and Outcome, ed. R.~{Capuzzo-Dolcetta}, M.~{Limongi}, \& A.~{Tornamb{\`e}}, 129

\bibitem[{{McMillan} {et~al.}(2012{\natexlab{b}}){McMillan}, {Portegies Zwart}, {van Elteren}, \& {Whitehead}}]{McMillan+2012}
{McMillan}, S., {Portegies Zwart}, S., {van Elteren}, A., \& {Whitehead}, A. 2012{\natexlab{b}}, in Astronomical Society of the Pacific Conference Series, Vol. 453, Advances in Computational Astrophysics: Methods, Tools, and Outcome, ed. R.~{Capuzzo-Dolcetta}, M.~{Limongi}, \& A.~{Tornamb{\`e}}, 129

\bibitem[{{Miyoshi} \& {Kusano}(2005)}]{2005JCoPh.208..315M}
{Miyoshi}, T. \& {Kusano}, K. 2005, Journal of Computational Physics, 208, 315

\bibitem[{{Motte} {et~al.}(2018){Motte}, {Bontemps}, \& {Louvet}}]{2018ARA&A..56...41M}
{Motte}, F., {Bontemps}, S., \& {Louvet}, F. 2018, \araa, 56, 41

\bibitem[{{Parmentier} \& {Pfalzner}(2013)}]{2013A&A...549A.132P}
{Parmentier}, G. \& {Pfalzner}, S. 2013, \aap, 549, A132

\bibitem[{{Pelupessy} {et~al.}(2013{\natexlab{a}}){Pelupessy}, {van Elteren}, {de Vries}, {McMillan}, {Drost}, \& {Portegies Zwart}}]{2013A&A...557A..84P}
{Pelupessy}, F.~I., {van Elteren}, A., {de Vries}, N., {et~al.} 2013{\natexlab{a}}, \aap, 557, A84

\bibitem[{{Pelupessy} {et~al.}(2013{\natexlab{b}}){Pelupessy}, {van Elteren}, {de Vries}, {McMillan}, {Drost}, \& {Portegies Zwart}}]{Pelupessy+2013}
{Pelupessy}, F.~I., {van Elteren}, A., {de Vries}, N., {et~al.} 2013{\natexlab{b}}, \aap, 557, A84

\bibitem[{{Polak} {et~al.}(2024){Polak}, {Mac Low}, {Klessen}, {Wei Teh}, {Cournoyer-Cloutier}, {Andersson}, {Appel}, {Tran}, {Lewis}, {Wilhelm}, {Portegies Zwart}, {Glover}, {Rieder}, {Wang}, \& {McMillan}}]{polak2024}
{Polak}, B., {Mac Low}, M.-M., {Klessen}, R.~S., {et~al.} 2024, \aap, 690, A94

\bibitem[{{Portegies Zwart} \& {McMillan}(2018)}]{PortegiesZwartMcMillan2018}
{Portegies Zwart}, S. \& {McMillan}, S. 2018, {Astrophysical Recipes: the art of AMUSE}, 2514-3433 (IOP Publishing)

\bibitem[{{Portegies Zwart} {et~al.}(2009){Portegies Zwart}, {McMillan}, {Harfst}, {Groen}, {Fujii}, {Nuall{\'a}in}, {Glebbeek}, {Heggie}, {Lombardi}, {Hut}, {Angelou}, {Banerjee}, {Belkus}, {Fragos}, {Fregeau}, {Gaburov}, {Izzard}, {Juri{\'c}}, {Justham}, {Sottoriva}, {Teuben}, {van Bever}, {Yaron}, \& {Zemp}}]{PortegiesZwart2009}
{Portegies Zwart}, S., {McMillan}, S., {Harfst}, S., {et~al.} 2009, \na, 14, 369

\bibitem[{{Portegies Zwart} {et~al.}(2013){Portegies Zwart}, {McMillan}, {van Elteren}, {Pelupessy}, \& {de Vries}}]{PortegiesZwart2013}
{Portegies Zwart}, S., {McMillan}, S.~L.~W., {van Elteren}, E., {Pelupessy}, I., \& {de Vries}, N. 2013, Computer Physics Communications, 184, 456

\bibitem[{{Portegies Zwart} {et~al.}(2020){Portegies Zwart}, {Pelupessy}, {Mart{\'\i}nez-Barbosa}, {van Elteren}, \& {McMillan}}]{2020CNSNS..8505240P}
{Portegies Zwart}, S., {Pelupessy}, I., {Mart{\'\i}nez-Barbosa}, C., {van Elteren}, A., \& {McMillan}, S. 2020, Communications in Nonlinear Science and Numerical Simulations, 85, 105240

\bibitem[{{Portegies Zwart} {et~al.}(2010){Portegies Zwart}, {McMillan}, \& {Gieles}}]{portegies-zwart2010}
{Portegies Zwart}, S.~F., {McMillan}, S.~L.~W., \& {Gieles}, M. 2010, \araa, 48, 431

\bibitem[{{Portegies Zwart} \& {Verbunt}(1996)}]{PortegiesZwartVerbunt1996}
{Portegies Zwart}, S.~F. \& {Verbunt}, F. 1996, \aap, 309, 179

\bibitem[{{Shukirgaliyev} {et~al.}(2021){Shukirgaliyev}, {Otebay}, {Sobolenko}, {Ishchenko}, {Borodina}, {Panamarev}, {Myrzakul}, {Kalambay}, {Naurzbayeva}, {Abdikamalov}, \& et~al.}]{Bek+2021}
{Shukirgaliyev}, B., {Otebay}, A., {Sobolenko}, M., {et~al.} 2021, \aap, 654, A53

\bibitem[{{Shukirgaliyev} {et~al.}(2017){Shukirgaliyev}, {Parmentier}, {Berczik}, \& {Just}}]{Bek+2017}
{Shukirgaliyev}, B., {Parmentier}, G., {Berczik}, P., \& {Just}, A. 2017, \aap, 605, A119

\bibitem[{{Shukirgaliyev} {et~al.}(2019){Shukirgaliyev}, {Parmentier}, {Berczik}, \& {Just}}]{bek+2019}
{Shukirgaliyev}, B., {Parmentier}, G., {Berczik}, P., \& {Just}, A. 2019, \mnras, 486, 1045

\bibitem[{{Shukirgaliyev} {et~al.}(2018){Shukirgaliyev}, {Parmentier}, {Just}, \& {Berczik}}]{bek+2018}
{Shukirgaliyev}, B., {Parmentier}, G., {Just}, A., \& {Berczik}, P. 2018, \apj, 863, 171

\bibitem[{{Skory} {et~al.}(2010){Skory}, {Turk}, \& {Norman}}]{2010ApJS..191...43S}
{Skory}, S., {Turk}, M.~J., \& {Norman}, M.~L. 2010, \apjs, 191, 43

\bibitem[{{Smith} {et~al.}(2009){Smith}, {Longmore}, \& {Bonnell}}]{2009MNRAS.400.1775S}
{Smith}, R.~J., {Longmore}, S., \& {Bonnell}, I. 2009, \mnras, 400, 1775

\bibitem[{{Sormani} {et~al.}(2017){Sormani}, {Tre{\ss}}, {Klessen}, \& {Glover}}]{2017MNRAS.466..407S}
{Sormani}, M.~C., {Tre{\ss}}, R.~G., {Klessen}, R.~S., \& {Glover}, S. C.~O. 2017, \mnras, 466, 407

\bibitem[{{Toonen} {et~al.}(2012{\natexlab{a}}){Toonen}, {Nelemans}, \& {Portegies Zwart}}]{2012A&A...546A..70T}
{Toonen}, S., {Nelemans}, G., \& {Portegies Zwart}, S. 2012{\natexlab{a}}, \aap, 546, A70

\bibitem[{{Toonen} {et~al.}(2012{\natexlab{b}}){Toonen}, {Nelemans}, \& {Portegies Zwart}}]{Toonen+2012}
{Toonen}, S., {Nelemans}, G., \& {Portegies Zwart}, S. 2012{\natexlab{b}}, \aap, 546, A70

\bibitem[{{Truelove} {et~al.}(1997){Truelove}, {Klein}, {McKee}, {Holliman}, {Howell}, \& {Greenough}}]{truelove1997}
{Truelove}, J.~K., {Klein}, R.~I., {McKee}, C.~F., {et~al.} 1997, \apjl, 489, L179

\bibitem[{{Turk} {et~al.}(2011){Turk}, {Smith}, {Oishi}, {Skory}, {Skillman}, {Abel}, \& {Norman}}]{Turk11a}
{Turk}, M.~J., {Smith}, B.~D., {Oishi}, J.~S., {et~al.} 2011, \apjs, 192, 9

\bibitem[{{Verliat} {et~al.}(2022){Verliat}, {Hennebelle}, {Gonz{\'a}lez}, {Lee}, \& {Geen}}]{2022A&A...663A...6V}
{Verliat}, A., {Hennebelle}, P., {Gonz{\'a}lez}, M., {Lee}, Y.-N., \& {Geen}, S. 2022, \aap, 663, A6

\bibitem[{Virtanen {et~al.}(2020)Virtanen, Gommers, Oliphant, Haberland, Reddy, Cournapeau, Burovski, Peterson, Weckesser, Bright, {van der Walt}, Brett, Wilson, Millman, Mayorov, Nelson, Jones, Kern, Larson, Carey, Polat, Feng, Moore, {VanderPlas}, Laxalde, Perktold, Cimrman, Henriksen, Quintero, Harris, Archibald, Ribeiro, Pedregosa, {van Mulbregt}, \& {SciPy 1.0 Contributors}}]{SciPy}
Virtanen, P., Gommers, R., Oliphant, T.~E., {et~al.} 2020, Nature Methods, 17, 261

\bibitem[{{Wall} {et~al.}(2020){Wall}, {Mac Low}, {McMillan}, {Klessen}, {Portegies Zwart}, \& {Pellegrino}}]{wall2020}
{Wall}, J.~E., {Mac Low}, M.-M., {McMillan}, S. L.~W., {et~al.} 2020, \apj, 904, 192

\bibitem[{{Wall} {et~al.}(2019){Wall}, {McMillan}, {Mac Low}, {Klessen}, \& {Portegies Zwart}}]{wall2019}
{Wall}, J.~E., {McMillan}, S. L.~W., {Mac Low}, M.-M., {Klessen}, R.~S., \& {Portegies Zwart}, S. 2019, \apj, 887, 62

\bibitem[{{W{\"u}nsch}(2015)}]{2015HiA....16..614W}
{W{\"u}nsch}, R. 2015, Highlights of Astronomy, 16, 614

\end{thebibliography}

\clearpage
\onecolumn
\appendix
\section{Observational comparison sample}
\label{app:obs_sample}

The observational reference sample used in Fig.~\ref{fig:q2d_obs} is listed in Table~\ref{tab:obs_q2d_sfincs}. It combines young stellar systems from the \textit{SFiNCs} project \citep{getman2017,getman2018} with the young open cluster sample analysed by \citet{dib2018}. We retain only systems younger than \(3.1\) Myr, matching the age range covered by the simulations being  compared.

For the \textit{SFiNCs} subsample, we adopt the published subcluster definitions, membership lists, stellar positions, and individual stellar ages. The source positions are converted from right ascension and declination to a local Cartesian projection, and \(Q_{\mathrm{2D}}\) is recomputed from the projected stellar distribution using the same minimum spanning tree and mean separation definition as in our simulation analysis. We estimate \(1\sigma\) uncertainties on \(Q_{\mathrm{2D}}\) using exact leave-one-out jackknife resampling over all members with valid positions. Median ages and their \(1\sigma\) jackknife uncertainties are computed analogously from the individual stellar ages.

We include only \textit{SFiNCs} systems with at least \(N_{\rm pos}\geq 20\) members with valid positions, in order to avoid cases where \(Q_{\mathrm{2D}}\) is dominated by a small number of stars. We do not impose an additional cut on the number of stars with age estimates. For systems with only a small number of age measurements, the median ages and their jackknife uncertainties should therefore be regarded as indicative rather than precise estimates. In this context, ``UnCl'' denotes the extended, spatially distributed unclustered component identified in the \textit{SFiNCs} mixture-model analysis, rather than one of the compact subcluster components \citep{getman2018}.

For the open cluster subsample, we adopt the published structural parameter \(Q_\mathrm{2D}\) values from \citet{dib2018}. The corresponding ages are taken from the MWSC catalogue \citep{mwsc1}. Since individual member positions, individual stellar ages, and uncertainties on \(Q_\mathrm{2D}\) are not available for this subsample, we do not recompute \(Q_{\mathrm{2D}}\) or estimate jackknife uncertainties. These systems are therefore treated as point estimates in the comparison.

\begin{table*}[t]
\caption{Observational comparison sample.} 
\label{tab:obs_q2d_sfincs}
\centering
\small
\setlength{\tabcolsep}{4pt}
\begin{tabular}{lcccccc@{\hspace{1.5em}}lcccccc}
\hline\hline
Cluster & Age & $\sigma_{\rm Age}$ & $Q_{\mathrm{2D}}$ & $\sigma_Q$ & $N_{\rm pos}$ & Ref. &
Cluster & Age & $\sigma_{\rm Age}$ & $Q_{\mathrm{2D}}$ & $\sigma_Q$ & $N_{\rm pos}$ & Ref. \\
 & [Myr] & [Myr] &  &  &  &  &
 & [Myr] & [Myr] &  &  &  &  \\
\hline
\multicolumn{14}{c}{\textit{SFiNCs subsample}} \\
\hline
Be 59 A            & 1.80 & 0.95 & 0.693 & 0.022 & 321 & SFiNCs &
Mon R2 B           & 1.50 & 0.38 & 0.603 & 0.037 & 127 & SFiNCs \\
Be 59 B            & 2.20 & 0.15 & 0.825 & 0.031 & 220 & SFiNCs &
Mon R2 C           & 0.80 & 0.14 & 0.828 & 0.077 &  32 & SFiNCs \\
Mon R2 UnCl        & 1.70 & 0.00 & 0.724 & 0.025 & 247 & SFiNCs &
GGD 12-15 A        & 0.65 & 0.17 & 0.787 & 0.049 & 108 & SFiNCs \\
NGC 1333 A         & 2.50 & 1.51 & 0.774 & 0.072 &  60 & SFiNCs &
GGD 12-15 UnCl     & 2.50 & 0.70 & 0.753 & 0.040 & 104 & SFiNCs \\
NGC 1333 B         & 1.70 & 0.23 & 0.654 & 0.048 & 101 & SFiNCs &
RCW 120 B          & 0.80 & 0.21 & 0.780 & 0.045 & 110 & SFiNCs \\
RCW 120 C          & 0.70 & 0.42 & 0.770 & 0.058 &  56 & SFiNCs &
IC 348 B           & 2.50 & 0.00 & 0.810 & 0.030 & 280 & SFiNCs \\
RCW 120 UnCl       & 1.25 & 0.40 & 0.735 & 0.029 & 157 & SFiNCs &
IC 348 A           & 2.30 & 0.22 & 0.767 & 0.073 &  28 & SFiNCs \\
Serpens Main B     & 0.60 & 0.27 & 0.713 & 0.054 &  61 & SFiNCs &
LkH$\alpha$ 101 A  & 1.45 & 0.36 & 0.843 & 0.035 & 182 & SFiNCs \\
Serpens Main UnCl  & 2.25 & 0.19 & 0.883 & 0.054 &  65 & SFiNCs &
LkH$\alpha$ 101 UnCl & 2.20 & 0.60 & 0.516 & 0.034 &  63 & SFiNCs \\
Serpens South UnCl & 1.80 & 0.77 & 0.698 & 0.028 & 199 & SFiNCs &
NGC 2068-2071 B    & 1.20 & 0.32 & 0.769 & 0.048 & 116 & SFiNCs \\
NGC 2068-2071 C    & 0.60 & 0.06 & 0.762 & 0.082 &  33 & SFiNCs &
IRAS 20050+2720 D  & 1.90 & 0.87 & 0.775 & 0.050 & 111 & SFiNCs \\
NGC 2068-2071 D    & 0.95 & 0.45 & 0.512 & 0.058 &  44 & SFiNCs &
NGC 2068-2071 UnCl & 2.30 & 0.59 & 0.461 & 0.026 & 117 & SFiNCs \\
IC 5146 B          & 1.55 & 0.28 & 0.784 & 0.041 & 142 & SFiNCs &
IC 5146 UnCl       & 2.60 & 0.59 & 0.675 & 0.046 &  93 & SFiNCs \\
ONC Flank S A      & 1.60 & 0.00 & 0.723 & 0.022 & 325 & SFiNCs &
ONC Flank N A      & 1.70 & 0.22 & 0.799 & 0.030 & 259 & SFiNCs \\
LDN 1251B UnCl     & 2.70 & 1.08 & 0.811 & 0.073 &  34 & SFiNCs &
OMC 2-3 A          & 0.90 & 0.00 & 0.599 & 0.065 &  54 & SFiNCs \\
Cep OB3b A         & 2.20 & 0.00 & 0.781 & 0.022 & 508 & SFiNCs &
Cep OB3b B         & 1.70 & 0.00 & 0.878 & 0.074 &  42 & SFiNCs \\
OMC 2-3 C          & 1.20 & 0.33 & 0.598 & 0.053 &  50 & SFiNCs &
Cep OB3b C         & 2.40 & 0.00 & 0.852 & 0.018 & 817 & SFiNCs \\
Cep A A            & 1.40 & 0.13 & 0.798 & 0.040 & 172 & SFiNCs &
OMC 2-3 UnCl       & 1.70 & 0.24 & 0.678 & 0.025 & 234 & SFiNCs \\
Cep A UnCl         & 2.00 & 0.00 & 0.638 & 0.035 &  98 & SFiNCs &
Mon R2 A           & 1.20 & 0.50 & 0.630 & 0.036 & 134 & SFiNCs \\
Cep C A            & 0.80 & 0.27 & 0.691 & 0.046 &  86 & SFiNCs &
Cep C UnCl         & 2.20 & 0.88 & 0.701 & 0.043 &  82 & SFiNCs \\
\hline
\multicolumn{14}{c}{\textit{Open-cluster subsample from Dib+18}} \\
\hline
Berkeley\_59       & 1.26 & --   & 0.834 & --    & --  & Dib+18 &
Cep\_OB4           & 1.26 & --   & 0.758 & --    & --  & Dib+18 \\
NGC\_1976          & 1.00 & --   & 0.680 & --    & --  & Dib+18 &
Collinder\_106     & 2.60 & --   & 0.780 & --    & --  & Dib+18 \\
Wit\_1             & 2.24 & --   & 0.793 & --    & --  & Dib+18 &
Pup\_OB3           & 2.51 & --   & 0.767 & --    & --  & Dib+18 \\
FSR\_1435          & 1.00 & --   & 0.815 & --    & --  & Dib+18 &
DBSB\_43           & 1.00 & --   & 0.867 & --    & --  & Dib+18 \\
Chamaleon\_I       & 1.00 & --   & 0.774 & --    & --  & Dib+18 &
Antalova\_2        & 1.00 & --   & 0.803 & --    & --  & Dib+18 \\
ASCC\_93           & 1.26 & --   & 0.751 & --    & --  & Dib+18 &
Sgr\_OB7           & 2.82 & --   & 0.741 & --    & --  & Dib+18 \\
NGC\_6611          & 2.14 & --   & 0.864 & --    & --  & Dib+18 &
NGC\_6618          & 1.00 & --   & 0.728 & --    & --  & Dib+18 \\
Sct\_OB3           & 1.26 & --   & 0.735 & --    & --  & Dib+18 &
Dolidze\_32        & 1.00 & --   & 0.781 & --    & --  & Dib+18 \\
Juchert\_3         & 1.00 & --   & 0.766 & --    & --  & Dib+18 &
FSR\_0148          & 1.00 & --   & 0.766 & --    & --  & Dib+18 \\
IC\_1396           & 1.00 & --   & 0.709 & --    & --  & Dib+18 &
ASCC\_117          & 1.68 & --   & 0.788 & --    & --  & Dib+18 \\
\hline
\end{tabular}
\tablefoot{Data used in Fig.~\ref{fig:q2d_obs}. For the \textit{SFiNCs} subsample, ages, memberships, and source positions are taken from \citet{getman2017,getman2018}; the variance in \(Q_{\mathrm{2D}}\), \(\sigma_Q\), and in median age, \(\sigma_{\rm Age}\), are computed using jackknife resampling. For the open-cluster subsample, \(Q_\mathrm{2D}\) values are taken from \citet{dib2018} and ages from the MWSC catalogue \citep{mwsc1}; no uncertainties or membership sizes are available, so \(\sigma_{\rm Age}\), \(\sigma_Q\), and \(N_{\rm pos}\) are not presented.}
\end{table*}

The black observational curve in Fig.~\ref{fig:q2d_obs} is constructed as an unweighted binned median of the combined \textit{SFiNCs} and \citet{dib2018} sample. The median is computed using only the central age and \(Q_{\mathrm{2D}}\) values. The jackknife uncertainties of the \textit{SFiNCs} systems are shown as error bars in the figure, but they are not used as weights because comparable uncertainties are not available for the \citet{dib2018} data.

\end{document}